\documentclass[12pt,a4paper]{article}
\usepackage{amsmath,amssymb,dsfont,graphicx,cite,subfigure}
\def\beq{\begin{equation}}
\def\eeq{\end{equation}}
\def\bea{\begin{eqnarray}}
\def\eea{\end{eqnarray}}
\newtheorem{theorem}{Theorem}
\newtheorem{prop}[theorem]{Proposition}

\newtheorem{definition}[theorem]{Definition}
\newtheorem{corollary}[theorem]{Corollary}
\newtheorem{conjecture}[theorem]{Conjecture}
\makeatletter
\expandafter\let\expandafter
\reset@font\csname reset@font\endcsname
\def\subeqnarray{\arraycolsep1pt
   \def\@eqnnum\stepcounter##1{\stepcounter{subequation}
       {\reset@font\rm(\theequation\alph{subequation})}}
\jot5mm     \eqnarray}

\makeatother

\newcommand{\bbZ}{{\mathbb Z}}
\newcommand{\bbR}{{\mathbb R}}
\newcommand{\bbQ}{{\mathbb Q}}

\def\ep{\epsilon}

\def\su2{{\mathfrak {su}}(2)}
\def\e3{{\mathfrak {e}}(3)}

\newbox\meibox
\def\placeunder#1#2#3#4{\setbox\meibox%
\vbox{\hbox{\hskip#4$\hphantom{#2}$}\hbox{$\hphantom{#1}$}}%
\vtop{\baselineskip=0pt\lineskiplimit=\baselineskip%
\lineskip=#3\hbox to \wd\meibox{\hfil\hskip#4$#2$\hfil}%
\hbox to \wd\meibox{\hfil$#1$\hfil}}}

\begin{document}
\title{On integrability of Hirota-Kimura type discretizations.
Experimental study of the discrete Clebsch system}

\author{Matteo Petrera\thanks{Dipartimento di Fisica,
Universit\`a degli Studi Roma Tre and Sezione INFN, Roma Tre, Via
della Vasca Navale 84, 00146 Roma, Italy. E-mail: {\tt
petrera}@{\tt fis.uniroma3.it}} \and Andreas
Pfadler\thanks{Zentrum Mathematik, Technische Universit\"at
M\"unchen, Boltzmannstr. 3, 85747 Garching bei M\"unchen, Germany.
E-Mail: {\tt pfadler}@{\tt in.tum.de}} \and Yuri B.
Suris\thanks{Zentrum Mathematik, Technische Universit\"at
M\"unchen, Boltzmannstr. 3, 85747 Garching bei M\"unchen, Germany.
E-Mail: {\tt suris}@{\tt ma.tum.de}}}

\maketitle

\begin{abstract}
R.~Hirota and K.~Kimura discovered integrable discretizations of
the Euler and the Lagrange tops, given by birational maps. Their
method is a specialization to the integrable context of a general
discretization scheme introduced by W.~Kahan and applicable to any
vector field with a quadratic dependence on phase variables.
According to a proposal by T.~Ratiu, discretizations of the
Hirota-Kimura type can be considered for numerous integrable
systems of classical mechanics. Due to a remarkable and not well
understood mechanism, such discretizations seem to inherit the
integrability for all algebraically completely integrable systems.
We introduce an experimental method for a rigorous study of
integrability of such discretizations. Application of this method
to the Hirota-Kimura type discretization of the Clebsch system
leads to the discovery of four functionally independent integrals
of motion of this discrete time system, which turn out to be much
more complicated than the integrals of the continuous time system.
Further, we prove that every orbit of the discrete time Clebsch
system lies in an intersection of four quadrics in the
six-dimensional phase space. Analogous results hold for the
Hirota-Kimura type discretizations for all commuting flows of the
Clebsch system, as well as for the $so(4)$ Euler top.
\end{abstract}

\section{Introduction}
\label{Sect: intro}

The discretization method studied in this paper seems to be
introduced in the geometric integration literature by W. Kahan in
the unpublished notes \cite{K}. It is applicable to any system of
ordinary differential equations for $x:\bbR\to\bbR^n$ with a
quadratic vector field:
\begin{equation}\label{eq: diff eq gen}
\dot{x}=Q(x)+Bx+c,
\end{equation}
where each component of $Q:\bbR^n\to\bbR^n$ is a quadratic form,
while $B\in{\rm Mat}_{n\times n}$ and $c\in\bbR^n$. Kahan's
discretization reads as
\begin{equation}\label{eq: Kahan gen}
\frac{\widetilde{x}-x}{2\epsilon}=Q(x,\widetilde{x})+
\frac{1}{2}B(x+\widetilde{x})+c,
\end{equation}
where
\[
Q(x,\widetilde{x})=\frac{1}{2}\Big(Q(x+\widetilde{x})-Q(x)-
Q(\widetilde{x})\Big)
\]
is the symmetric bilinear form corresponding to the quadratic form
$Q$. Here and below we use the following notational convention
which will allow us to omit a lot of indices: for a sequence
$x:\bbZ\to\bbR$ we write $x$ for $x_k$ and $\widetilde{x}$ for
$x_{k+1}$. Eq. (\ref{eq: Kahan gen}) is {\em linear} with respect
to $\widetilde x$ and therefore defines a {\em rational} map
$\widetilde{x}=f(x,\epsilon)$. Clearly, this map approximates the
time-$(2\epsilon)$-shift along the solutions of the original
differential system, so that $x_k\approx x(2k\epsilon)$. (We have
chosen a slightly unusual notation $2\epsilon$ for the time step,
in order to avoid appearance of various powers of 2 in numerous
formulas; a more standard choice would lead to changing
$\epsilon\mapsto\epsilon/2$ everywhere.) Since eq. (\ref{eq: Kahan
gen}) remains invariant under the interchange
$x\leftrightarrow\widetilde{x}$ with the simultaneous sign
inversion $\epsilon\mapsto-\epsilon$, one has the {\em
reversibility} property
\begin{equation}\label{eq: reversible}
f^{-1}(x,\epsilon)=f(x,-\epsilon).
\end{equation}
In particular, the map $f$ is {\em birational}.

W.~Kahan applied this discretization scheme to the famous
Lotka-Volterra system and showed that in this case it possesses a
very remarkable non-spiralling property. We will briefly discuss
this example in Sect. \ref{Sect: LV}. Some further applications of
this discretization have been explored in \cite{KHLI}.

The next, even more intriguing appearance of this discretization
was in the two papers by R. Hirota and K. Kimura who (being
apparently unaware of the work by Kahan) applied it to two famous
{\em integrable} system of classical mechanics, the Euler top and
the Lagrange top \cite{HK, KH}. For the purposes of the present
text, integrability of a dynamical system is synonymous with the
existence of a sufficient number of functionally independent
conserved quantities, or integrals of motion, that is, functions
constant along the orbits. We leave aside other aspects of the
multi-facet notion of integrability, such as Hamiltonian ones or
explicit solution. Surprisingly, the Kahan-Hirota-Kimura
discretization scheme produced in both the Euler and the Lagrange
cases of the rigid body motion {\em integrable} maps. Even more
surprisingly, the mechanism which assures integrability in these
two cases seems to be rather different from the majority of
examples known in the area of integrable discretizations, and,
more generally, integrable maps, cf. \cite{S}. The case of the
discrete time Euler top is relatively simple, and the proof of its
integrability given in \cite{HK} is rather straightforward and
easy to verify by hands. As it often happens, no explanation was
given in \cite{HK} about how this result has been {\em
discovered}. The ``derivation'' of integrals of motion for the
discrete time Lagrange top in \cite{KH} is rather cryptic and
almost uncomprehensible.

The present paper aims at clarifying the Hirota-Kimura
integrability mechanism and at its application to further
integrable systems. We use the term ``Hirota-Kimura type
discretization'' for the Kahan's discretization in the context of
integrable systems. In Sect. \ref{Sect: HK mechanism} we propose a
formalization of the Hirota-Kimura mechanism from \cite{KH}, which
will hopefully unveil its main idea and contribute towards
demystifying at least some of its aspects. We introduce a notion
of a ``Hirota-Kimura basis'' for a given map $f$. Such a basis
$\Phi$ is a set of simple (often monomial) functions,
$\Phi=(\varphi_l,\ldots,\varphi_l)$, such that for every orbit
$\{f^i(x)\}$ of the map $f$ there is a certain linear combination
$c_1\varphi_1+\ldots+c_l\varphi_l$ of functions from $\Phi$
vanishing on this orbit. As explained in Sect. \ref{Sect: HK
mechanism}, this is a {\em new} mathematical notion, not reducible
to that of integrals of motion, although closely related to the
latter. In Sect. \ref{Sect: finding} we lay a theoretical
fundament for the search for Hirota-Kimura bases for a given
discrete time system, and give a number of practical recipes and
tricks for doing this.

We dare to claim that the results of \cite{HK} concerning the
discrete time Euler top were originally discovered using the
mechanism of Hirota-Kimura bases, and we present in Sect.
\ref{Sect: dET} an attempt to reconstruct the way this discovery
has been made. Sect. \ref{Sect: Clebsch1} contains the main
results of this paper, namely the proof of integrability of the
Hirota-Kimura type discretization for a further famous integrable
system of the classical mechanics, namely for the Clebsch case of
the motion of a rigid body in an ideal fluid.

Our investigations are based mainly on computer experiments, which
are used both for {\em discovery} of new results and for their
rigorous {\em proof}. A search for Hirota-Kimura bases can be done
with the help of {\em numerical experiments} based on the recipe
(N) formulated in Sect. \ref{Sect: finding}, which has a
theoretical justification in Theorem \ref{Lem: dim K}. If the
search has been successful and a certain set of functions $\Phi$
has been identified as a Hirota-Kimura basis for a given map $f$,
then numerical experiments can provide a very convincing evidence
in favor of such a statement. A rigorous proof of such a statement
turns out to be much more demanding. At present, we are not in
possession of any theoretical proof strategies and are forced to
verify the corresponding statements by means of {\em symbolic
computations}. However, direct and simple-minded symbolic
computations turn out to be non-feasible due to complexity issues.
As detailed in Sect. \ref{Sect: Clebsch1}, the sheer size of
explicit expressions for the second iterate $f^2$ of the discrete
time Clebsch system precludes symbolic manipulations, like
solution of linear systems, as soon as those involve $f^2$.
Therefore our main effort has been put into finding the strategy
of a complete and rigorous symbolical proof which would avoid
using $f^2$ and would stay within the memory and performance
restrictions of the available software and hardware. The resulting
proofs are {\em computer assisted} and are based on symbolic
computations with MAPLE \cite{maple}, SINGULAR \cite{singular} and
FORM \cite{form}.

Our work was stimulated by a talk by T.~Ratiu at the Oberwolfach
Workshop ``Geometric Integration'' \cite{RC}, where an extension
of the Hirota-Kimura approach to the Clebsch system and to the
Kovalevski top has been proposed. However, no valid derivation of
integrals was presented in T.~Ratiu's talk, so that the question
on the integrability of these discretizations remained open. Our
work answers this question in the affirmative for the Clebsch
system (actually, even for a whole family of Hamiltonian flows
generated by commuting integrals of the Clebsch system). In the
concluding Sect. \ref{Sect: conclusions}, we discuss further
perspectives of this approach and formulate a general conjecture
about the integrability of the Hirota-Kimura type
discretizations..

\section{Kahan's discretization of the Lotka-Volterra system}
\label{Sect: LV}

\begin{figure}\label{Fig: LV}
\begin{center}
\includegraphics[width=0.5\textwidth]{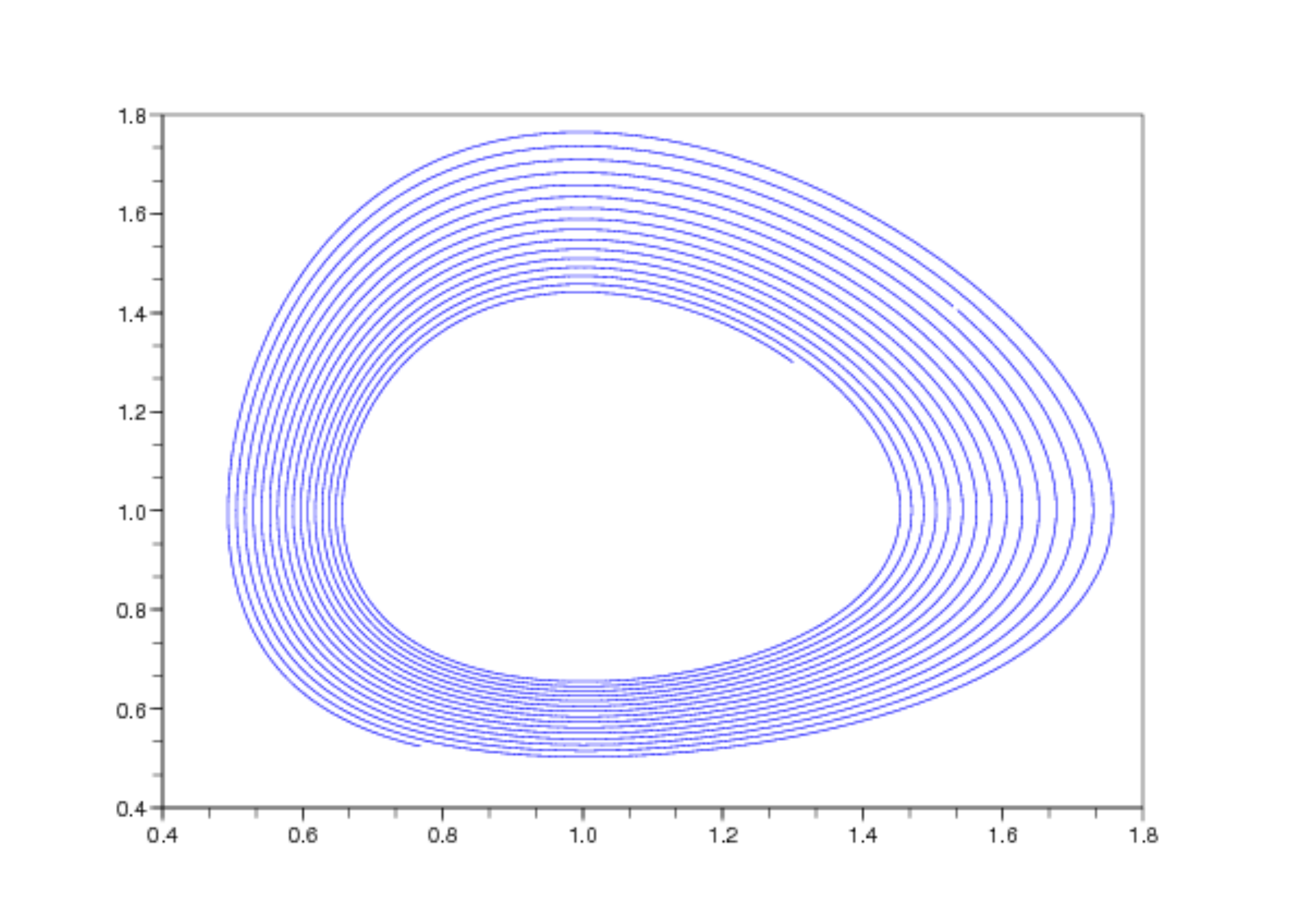}
\includegraphics[width=0.45\textwidth]{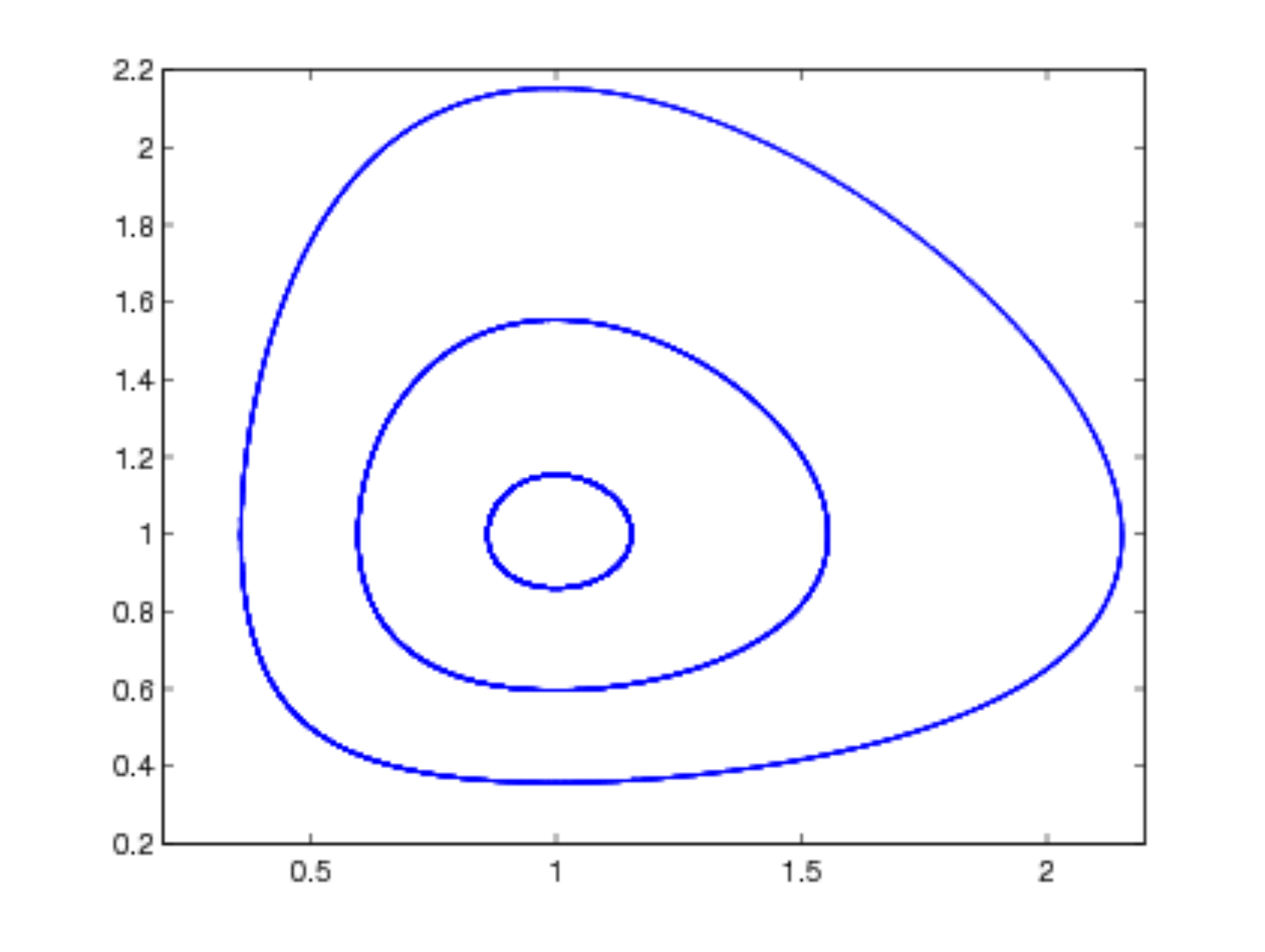}
\end{center}
\caption{Left: a spiralling orbit of the explicit Euler method
with the time-step $\epsilon=0.01$ applied to the Lotka-Volterra
system. Right: three orbits of the Kahan's discretization with
$\epsilon=0.1$.}
\end{figure}

As already mentioned in Sect. \ref{Sect: intro}, W.~Kahan applied
his general discretization scheme to the famous Lotka-Volterra
system modelling the interaction of the predator and the prey
populations:
\begin{equation}\label{eq: LV}
\dot{x}=x(1-y),\qquad \dot{y}=y(x-1).
\end{equation}
Solutions of this system lie on closed curves in (the first
quadrant of) the phase plane $\bbR^2$, because of the presence of
the integral (conserved quantity)
\begin{equation}
\nonumber
H(x,y)=x+y-\log(xy).
\end{equation}
Actually, system (\ref{eq: LV}) is Hamiltonian with respect to the
Poisson bracket
\begin{equation}
\label{eq: LV PB} \{x,y\}=xy,
\end{equation}
with the Hamilton function $H$:
\[
\dot{x}=-xy\frac{\partial H}{\partial y}\,,\qquad
\dot{y}=xy\frac{\partial H}{\partial x}\,.
\]
The majority of the conventional discretization schemes produce,
when applied to (\ref{eq: LV}), spiralling solutions. Compared
with solutions of the original system, this is a qualitatively
different behavior, cf. Fig. \ref{Fig: LV} (left). The
discretization proposed by Kahan reads:
\begin{equation}\label{eq: Kahan LV}
(\widetilde{x}-x)/\epsilon=(\widetilde{x}+x)-(\widetilde{x}y+x\widetilde{y}),\qquad
(\widetilde{y}-y)/\epsilon=(\widetilde{x}y+x\widetilde{y})-(\widetilde{y}+y),
\end{equation}
Eq. (\ref{eq: Kahan LV}) can be written as a linear system for
$(\widetilde{x},\widetilde{y})$,
\begin{equation}
\nonumber
\begin{pmatrix} 1-\epsilon+\epsilon y & \epsilon x\\
  -\epsilon y & 1+\epsilon-\epsilon x \end{pmatrix}
\begin{pmatrix}\widetilde{x} \\ \widetilde{y}\end{pmatrix}
=\begin{pmatrix} (1+\epsilon)x \\ (1-\epsilon)y\end{pmatrix},
\end{equation}
which can be immediately solved, thus yielding an {\em explicit}
map $(\widetilde x,\widetilde y)=f(x,y,\epsilon)$:
\begin{equation}\label{eq: Kahan LV explicit}
\left\{ \begin{array}{l}
\widetilde{x}=x\,\dfrac{(1+\epsilon)^2-\epsilon(1+\epsilon)x-\epsilon(1-\epsilon)y}
{1-\epsilon^2-\epsilon(1-\epsilon)x+\epsilon(1+\epsilon)y},\\
\widetilde{y}=y\,\dfrac{(1-\epsilon)^2+\epsilon(1+\epsilon)x+\epsilon(1-\epsilon)y}
{1-\epsilon^2-\epsilon(1-\epsilon)x+\epsilon(1+\epsilon)y}.
   \end{array}\right.
\end{equation}
A remarkable property of the Kahan's discretization is that it
apparently does not suffer from spiralling, solutions seem to fill
out closed curves in the phase plane, cf. Fig. \ref{Fig: LV}
(right). A (partial) explanation of this behavior was given in
\cite{SZ}, where it was shown that the map $f$ is Poisson with
respect to the invariant Poisson bracket (\ref{eq: LV PB}) of the
system (\ref{eq: LV}). It is unknown whether the map \ref{eq:
Kahan LV explicit} possesses an integral of motion, thus forcing
all orbits to lie on smooth closed curves, as suggested by Fig.
\ref{Fig: LV} (right). Some numerical experiments, via a deep
zoom-in into certain domains of the phase plane, indicate that the
map might be non-integrable, but a rigorous proof of a
non-existence statement seems to be rather difficult. It might be
possible with the use of technology described in \cite{GL}.

\section{Hirota-Kimura bases and integrals}
\label{Sect: HK mechanism}

In this section a general formulation of a remarkable mechanism
will be given, which seems to be responsible for the integrability
of the Hirota-Kimura type (or Kahan type) discretizations of
algebraically completely integrable systems. This mechanism is so
far not well understood, in fact at the moment we do not know what
mathematical structures make it actually work.

Throughout this section $f:\bbR^n\to\bbR^n$ is a birational map,
while $h_i,\varphi_i:\bbR^n\to\bbR$ stand for rational, usually
polynomial functions on the phase space. We start with recalling a
well known definition.

\begin{definition}\label{Def: integral}
A function $h:\bbR^n\to\bbR$ is called an {\bf integral}, or a
{\bf conserved quantity}, of the map $f$, if for every
$x_0\in\bbR^n$ there holds
\begin{equation}
\nonumber
h(f(x))=h(x),
\end{equation}
so that
\begin{equation}
\nonumber
h\circ f^i(x)=h(x)\quad \forall i\in\bbZ.
\end{equation}
\end{definition}

\noindent{\bf Convention.} In the last formula and everywhere in
the sequel, we use the expression $h\circ f^i(x)$ for the
evaluation of the function $h\circ f^i$ at the point $x$. This is
equivalent to $h(f^i(x))$ and is used to spare some parentheses.
\medskip

Thus, each orbit of the map $f$ lies on a certain level set of its
integral $h$. As a consequence, if one knows $d$ functionally
independent integrals $h_1,\ldots,h_d$ of $f$, one can claim that
each orbit of $f$ is confined to an $(n-d)$-dimensional invariant
set, which is a common level set of the functions
$h_1,\ldots,h_d$.

\begin{definition}\label{Def: Hirota mech}
A set of functions $\Phi=(\varphi_1,\ldots,\varphi_l)$, linearly
independent over $\bbR$, is called a {\bf Hirota-Kimura basis
(HK-basis)}, if for every $x_0\in\bbR^n$ there exists a vector
$c=(c_1,\ldots,c_l)\neq 0$ such that
\begin{equation}\label{eq: fundamental}
(c_1\varphi_1+\ldots+c_l\varphi_l)\circ f^i(x)=0\quad \forall
i\in\bbZ.
\end{equation}
For a given $x\in\bbR^n$, the set of all vectors $c\in\bbR^l$ with
this property will be denoted by $K_\Phi(x)$ and called the
null-space of the basis $\Phi$ (at the point $x$). This set
clearly is a vector space.
\end{definition}

Thus, for a HK-basis $\Phi$ and for $c\in K_\Phi(x)$ the function
$h=c_1\varphi_1 + ... + c_l\varphi_l$ vanishes along the $f$-orbit
of $x$. Let us stress that we cannot claim that $h=c_1\varphi_1 +
... + c_l\varphi_l$ is an integral of motion, since vectors $c\in
K_\Phi(x)$ do not have to belong to $K_\Phi(y)$ for initial points
$y$ not lying on the orbit of $x$. However, for any $x$ the orbit
$\{f^i(x)\}$ is confined to the common zero level set of $d$
functions
\begin{equation}
\nonumber
h_j=c_1^{(j)}\varphi_1+\ldots+c_l^{(j)}\varphi_l=0,\quad
j=1,\ldots,d,
\end{equation}
where the vectors
$c^{(j)}=\big(c_1^{(j)},\ldots,c_l^{(j)}\big)\in\bbR^l$ form a
basis of $K_\Phi(x)$. Thus, knowledge of a HK-basis with the
null-space of dimension $d$ leads to a similar conclusion as
knowledge of $d$ independent integrals of $f$, namely to the
conclusion that the orbits lie on $(n-d)$-dimensional invariant
sets. Note, however, that a HK-basis gives no immediate
information on how these invariant sets foliate the phase space
$\bbR^n$, since the vectors $c^{(j)}$, and therefore the functions
$h_j$, change from one initial point $x$ to another.

Although the notions of integrals and of HK-bases cannot be
immediately translated into one another, they turn out to be
closely related.

The simplest situation for a HK-basis corresponds to $l=2$, $\dim
K_\Phi(x)=d=1$. In this case we immediately see that
$h=\varphi_1/\varphi_2$ is an integral of motion of the map $f$.
Conversely, for any rational integral of motion
$h=\varphi_1/\varphi_2$ its numerator and denominator $\varphi_1$,
$\varphi_2$ satisfy
\[
(c_1\varphi_1+c_2\varphi_2)\circ f^i(x)=0,\quad i\in\bbZ,
\]
with $c_1=1$, $c_2=-h(x)$, and thus build a HK-basis with $l=2$.
Thus, the notion of a HK-basis {\em generalizes} (for $l\ge 3$)
the notion of integrals of motion.

On the other hand, knowing a HK-basis $\Phi$ with $\dim
K_\Phi(x)=d\ge 1$ allows one to find integrals of motion for the
map $f$. Indeed, from Definition \ref{Def: Hirota mech} there
follows immediately:
\begin{prop}\label{Th: K_integral}
If $\Phi$ is a HK-basis for a map $f$, then
\begin{equation}
\nonumber
K_\Phi(f(x))=K_\Phi(x).
\end{equation}
\end{prop}
Thus, the $d$-dimensional null-space $K_\Phi(x)\in Gr(d,l)$,
regarded as a function of the initial point $x\in\bbR^n$, is
constant along trajectories of the map $f$, i.e., it is a
$Gr(d,l)$-valued integral. One can extract from this fact a number
of scalar integrals.
\begin{corollary}\label{Th: mechanism} Let $\Phi$ be a
HK-basis for $f$ with $\dim K_\Phi(x)=d$ for all
$x\in\mathbb{R}^n$. Take a basis of $K_\Phi(x)$ consisting of $d$
vectors $c^{(i)}\in\bbR^l$ and put them into the columns of a
$l\times d$ matrix $C(x)$. For any $d$-index
$\alpha=(\alpha_1,\ldots,\alpha_d)\subset\{1,2,\ldots,n\}$ let
$C_\alpha=C_{\alpha_1\ldots\alpha_d}$ denote the $d\times d$ minor
of the matrix $C$ built from the rows $\alpha_1,\ldots,\alpha_d$.
Then for any two $d$-indices $\alpha,\beta$ the function
$C_\alpha/C_\beta$ is an integral of $f$.
\end{corollary}
{\bf Proof.} The functions $C_\alpha$ are nothing other than the
Grassmann-Pl\"ucker coordinates of the $d$-space $K_\Phi(x)$ in
the Grassmannian $Gr(d,l)$, which are defined up to a common
factor. More detailed, any basis of $K_\Phi(f(x))$ is obtained
from the given basis of $K_\Phi(x)$ via a right multiplication of
$C$ by a non-degenerate $d\times d$ matrix $D$. This yields a
simultaneous multiplication of all $C_\alpha$ by the common factor
$\det D$. This operation does not change the quotients
$C_\alpha/C_\beta$. \hfill $\Box$

Especially simple is the situation when the null-space of a
HK-basis has dimension $d=1$.
\begin{corollary}\label{Th: mechanism d=1} Let $\Phi$ be a
HK-basis for $f$ with $\dim K_\Phi(x)=1$ for all
$x\in\mathbb{R}^n$. Let $K_\Phi(x)=[c_1(x):\ldots :c_l(x)]\in
\bbR\mathbb{P}^{l-1}$. Then the functions $c_j/c_k$ are integrals
of motion for $f$.
\end{corollary}
An interesting (and difficult) question is about the number of
functionally independent integrals obtained from a given HK-basis
according to Corollaries \ref{Th: mechanism} and \ref{Th:
mechanism d=1}. We will see later that it is possible for a
HK-basis with a one-dimensional null-space to produce more than
one independent integral (see Theorem \ref{Th: dC basis 2}).

The first examples of this mechanism (with $d=1$) were found in
\cite{KH} and (somewhat implicitly) in \cite{HK}.

\section{Finding Hirota-Kimura bases}
\label{Sect: finding}

At present, we cannot give any theoretical sufficient conditions
for existence of a Hirota-Kimura basis $\Phi$ for a given map $f$,
and the only way to find such a basis remains the experimental
one. Definition \ref{Def: Hirota mech} requires to verify
condition (\ref{eq: fundamental}) for all $i\in\bbZ$, which is, of
course, impractical. We now show that it is enough to check this
condition for a finite number of iterates $f^i$.

For a given set of functions $\Phi=(\varphi_1,\ldots,\varphi_l)$
and for any interval $[j,k]\subset\bbZ$ we denote
\begin{equation}\label{eq: balance}
X_{[j,k]}(x) = \left(
\begin{array}{ccc}
\varphi_1(f^j(x))  & ..   & \varphi_l(f^j(x))   \\
\varphi_1(f^{j+1}(x))  & ..   & \varphi_l(f^{j+1}(x))   \\
...  &   & ...   \\
\varphi_1(f^{k}(x))  & .. & \varphi_l(f^{k}(x))
\end{array}
\right).
\end{equation}
In particular, $X_{(-\infty,\infty)}(x)$ will denote the double
infinite matrix of the type (\ref{eq: balance}). Obviously,
\[
\ker X_{(-\infty,\infty)}(x)=K_\Phi(x).
\]
Thus, Definition \ref{Def: Hirota mech} requires that $\dim \ker
X_{(-\infty,\infty)}(x)\ge 1$. Our algorithm for detecting this
situation is based on the following observation.
\begin{theorem}\label{Lem: dim K}
Let
\begin{equation}\label{eq: fundamental finite}
\dim\ker X_{[0,s-1]}(x)=\left\{\begin{array}{cl} l-s & {\rm
for}\;\; 1 \leq s \leq l-d, \\ d & {\rm for}\;\; s=l-d+1,
\end{array}\right.
\end{equation}
hold with some $d$ for all $x\in\bbR^n$. Then for any $x\in\bbR^n$
there holds:
\begin{equation}\nonumber
\ker X_{(-\infty,\infty)}(x)=\ker X_{[0,l-d-1]}(x),
\end{equation}
and, in particular,
\begin{equation}\nonumber
\dim\ker X_{(-\infty,\infty)}(x)=d.
\end{equation}
\end{theorem}
{\bf Proof.} By definition, $X_{[j,k]}(x)=X_{[0,k-j]}(f^j(x))$.
Therefore, applying condition (\ref{eq: fundamental finite}) to
iterates $f^j(x)$ instead of $x$ itself, we see that the kernel of
any submatrix of $X_{(-\infty,\infty)}(x)$ with $l-d$ rows, as
well as the kernel of any submatrix with $l-d+1$ rows, is
$d$-dimensional:
\[
\dim\ker X_{[j,j+l-d-1]}(x)=\dim\ker X_{[j,j+l-d]}(x)= \dim\ker
X_{[j+1,j+l-d]}(x).
\]
Since, obviously,
\[
\ker X_{[j,j+l-d-1]}(x)\supset\ker X_{[j,j+l-d]}(x)\subset\ker
X_{[j+1,j+l-d]}(x),
\]
we find that all three kernels coincide, in particular,
\[
\ker X_{[j,j+l-d-1]}(x)=\ker X_{[j+1,j+l-d]}(x).
\]
By induction, all $\ker X_{[j,j+l-d-1]}(x)$, $j\in\bbZ$, coincide,
and therefore they coincide with $\ker X_{(-\infty,\infty)}(x)$,
as well. \hfill $\Box$

These results lead us to formulate the following {\em numerical
algorithm for the estimation of $\dim K_\Phi(x)$ for a hypothetic
HK-basis} $\Phi=(\varphi_1,\ldots,\varphi_l)$.
\begin{itemize}
\item[(N)] For several randomly chosen initial points $x\in
\bbR^n$, compute $\dim\ker X_{[0,s-1]}(x)$ for $1 \leq s \leq l$.
If for every $x$ condition (\ref{eq: fundamental finite}) is
satisfied with one and the same $d\ge 1$, then $\Phi$ is likely to
be a HK-basis for $f$, with $\dim K_\Phi(x)=d$.
\end{itemize}
We stress once again that generally (for general maps $f$ and
general monomial sets $\Phi$) one will find that the $l\times l$
matrix $X_{[0,l-1]}(x)$ is non-degenerate for a typical $x$, so
that $\dim K_\Phi(x)=0$. Finding (a candidate for) a HK-basis
$\Phi$ is a highly non-trivial task.

Having found a HK-basis $\Phi$ with $\dim K_\Phi(x)=d$
numerically, one faces the next problem: to {\em prove} this fact,
that is, to prove that the system of equations (\ref{eq:
fundamental}) with $i=i_0,i_0+1,\ldots,i_0+l-d$ admits (for some,
and then for all $i_0\in\bbZ$) a $d$-dimensional space of
solutions. For the sake of clarity, we restrict our following
discussion to the most important case $d=1$. Thus, one has to
prove that the homogeneous system
\begin{equation}\label{eq: bigsystem}
(c_1 \varphi_1 + \ldots +c_l \varphi_l)\circ f^i(x)= 0 ,\qquad
i=i_0,i_0+1,\ldots,i_0+l-1
\end{equation}
admits for every $x\in\bbR^n$ a one-dimensional vector space of
non-trivial solutions. The main obstruction for a symbolic
solution of the system (\ref{eq: bigsystem}) is the growing
complexity of the iterates $f^i(x)$. While the expression for
$f(x)$ is typically of a moderate size, already the second iterate
$f^2(x)$ becomes typically prohibitively big. In such a situation
a symbolic solution of the linear system (\ref{eq: bigsystem})
should be considered as impossible, as soon as $f^2(x)$ is
involved, for instance, if $l\ge 3$ and one considers the linear
system with $i=0,1,\ldots,l-1$.

Therefore it becomes crucial to reduce the number of iterates
involved in (\ref{eq: bigsystem}) as far as possible. A reduction
of this number by 1 becomes in many cases crucial! One can imagine
several ways to accomplish this.

\begin{itemize}
\item[(A)] Take into account that, because of the reversibility
$f^{-1}(x,\epsilon)=f(x,-\epsilon)$, the negative iterates
$f^{-i}$ are of the same complexity as $f^i$. Therefore, one can
reduce the complexity of the functions involved in (\ref{eq:
bigsystem}) by choosing $i_0=-[l/2]$ instead of the naive choice
$i_0=0$.
\end{itemize}
For instance, in the case $l=3$ one should consider the system
(\ref{eq: bigsystem}) with $i=-1,0,1$, and not with $i=0,1,2$.
However, already in the case $l=4$ this simple recipe does not
allow us to avoid considering $f^2$. In this case, the following
way of dealing with the system (\ref{eq: bigsystem}) becomes
useful.

\begin{itemize}
\item[(B)] Set $c_l=-1$ and consider instead of the homogeneous
system (\ref{eq: bigsystem}) of $l$ equations the non-homogeneous
system
\begin{equation}\label{eq: smaller system}
(c_1 \varphi_1 +\ldots +c_{l-1} \varphi_{l-1})\circ f^i(x)=
\varphi_l\circ f^i(x),\quad i=i_0,i_0+1,\ldots,i_0+l-2,\quad
\end{equation}
of $l-1$ equations. Having found the (unique) solution
$\big(c_1(x),\ldots,c_{l-1}(x)\big)$, prove that these functions
are integrals of motion, that is,
\begin{equation}\label{eq: integrals}
c_1(f(x))=c_1(x),\quad\ldots,\quad c_{l-1}(f(x))=c_{l-1}(x).
\end{equation}
\end{itemize}
Thus, for instance, in the case $l=4$ one has to deal with the
non-homogeneous system of equations (\ref{eq: smaller system})
with $i=-1,0,1$. Unfortunately, even if one is able to solve this
system symbolically, the task of a symbolic verification of eq.
(\ref{eq: integrals}) might become very hard due to complexity of
the solutions $\big(c_1(x),\ldots,c_{l-1}(x)\big)$.

This is the way taken, for instance, in \cite{KH}. In that paper,
the task of verifying the equations of the type \eqref{eq:
integrals} for the discrete time Lagrange top is performed with
the following method.

\begin{itemize}
\item[(G)] In order to verify that a rational function
$c(x)=p(x)/q(x)$ is an integral of motion of the map $\widetilde
x=f(x)$ coming from a system \eqref{eq: Kahan gen}:
\begin{itemize}
 \item [i)] find a Gr\"obner basis $G$ of the ideal $I$ generated by the
components of eq. \eqref{eq: Kahan gen}, considered as
multi-linear polynomials of $2n$ variables $x,\widetilde x$ of
total degree 2;
 \item[ii)] check, via polynomial division through elements of $G$, whether
 the polynomial $\delta(x,\widetilde x)=p(\widetilde x)q(x)-p(x)q(\widetilde x)$
 belongs to the ideal $I$.
\end{itemize}
\end{itemize}
An advantage of this method is that neither of its two steps needs
the complicated explicit expressions for the map $f$.
Nevertheless, both steps might be very demanding, especially the
second step in case of a complicated integral $c(x)$.

Sometimes, the task of verifying equations \eqref{eq: integrals}
can be circumvented  by means of the following tricks.

\begin{itemize}
\item[(C)] Solve system (\ref{eq: smaller system}) for two
different but overlapping ranges $i\in[i_0,i_0+l-2]$ and
$i\in[i_1,i_1+l-2]$. If the solutions coincide, then eq. (\ref{eq:
integrals}) holds automatically.
\end{itemize}
Indeed, in this situation the functions
$\big(c_1(x),\ldots,c_{l-1}(x)\big)$ solve the system with $i\in
[i_0,i_0+l-2]\cup[i_1,i_1+l-2]$ consisting of more than $l-1$
equations.

A clever modification of this idea, which allows one to avoid
solving the second system, is as follows.
\begin{itemize}
\item[(D)] Suppose that the index range $i\in[i_0,i_0+l-2]$ in eq.
(\ref{eq: smaller system}) contains 0 but is non-symmetric. If the
solution of this system
$\big(c_1(x,\epsilon),\ldots,c_{l-1}(x,\epsilon)\big)$ is even
with respect to $\epsilon$, then eqs. (\ref{eq: integrals}) hold
automatically.
\end{itemize}
Indeed, the reversibility of the map
$f^{-1}(x,\epsilon)=f(x,-\epsilon)$ yields in this case that
equations of the system (\ref{eq: smaller system}) are satisfied
for $i\in[-(i_0+l-2),-i_0]$, as well, and the intervals
$[i_0,i_0+l-2]$ and $[-(i_0+l-2),-i_0]$ overlap but do not
coincide, by condition.

Finally, the most powerful method of reducing the number of
iterations to be considered is as follows.
\begin{itemize}
\item[(E)] Often, the solutions
$\big(c_1(x),\ldots,c_{l-1}(x)\big)$ satisfy some linear relations
with constant coefficients. Find (observe) such relations {\em
numerically}. Each such (still hypothetic) relation can be used to
replace one equation in the system (\ref{eq: smaller system}).
Solve the resulting system {\em symbolically}, and proceed as in
recipes (C) or (D) in order to verify eqs. (\ref{eq: integrals}).
\end{itemize}

In some (rare) cases the integrals found by this approach are nice
and simple enough to enable one to verify eqs. (\ref{eq:
integrals}) directly. Of course, it would be highly desirable to
find some structures, like Lax representation, bi-Hamiltonian
structure, etc., which would allow one to check the conservation
of integrals in a more clever way, but up to now no such
structures have been found for any of the Hirota-Kimura-type
discretizations.

\section{Hirota-Kimura discretization of the Euler top}
\label{Sect: dET}

We now illustrate the Hirota-Kimura mechanism by its application
to the Euler top. This three-dimensional system is simple enough
to enable one to perform all necessary computations symbolically,
even by hand. At the same time, it provides a perfect illustration
for many of the issues mentioned in the previous section.

\subsection{Euler top}
The differential equations of motion of the Euler top read
\begin{equation}\label{eq: ET x}
\dot{x}_1=\alpha_1 x_2 x_3,\quad \dot{x}_2=\alpha_2 x_3 x_1, \quad
\dot{x}_3=\alpha_3 x_1x_2,
\end{equation}
with $\alpha_i$ being real parameters of the system. This is one
of the most famous integrable systems of the classical mechanics,
with a big literature devoted to it. We mention only that this
system can be explicitly integrated in terms of elliptic
functions, and admits two functionally independent integrals of
motion. Actually, a quadratic function
$H(x)=\gamma_1x_1^2+\gamma_2x_2^2+\gamma_3x_3^2$ is an integral
for eqs. (\ref{eq: ET x}), if $\langle
\gamma,\alpha\rangle=\gamma_1\alpha_1+\gamma_2\alpha_2+\gamma_2\alpha_2=0$.
In particular, the following three functions are integrals of
motion:
\begin{equation}\nonumber
H_1=\alpha_3x_2^2-\alpha_2x_3^2,\qquad
H_2=\alpha_1x_3^2-\alpha_3x_1^2,\qquad
H_3=\alpha_2x_1^2-\alpha_1x_2^2.
\end{equation}
Clearly, only two of them are functionally independent because of
$\alpha_1H_1+\alpha_2H_2+\alpha_3H_3=0$.

\subsection{Discrete equations of motion}
\begin{figure}\label{Fig: dET}
\begin{center}
\includegraphics[width=0.45\textwidth]{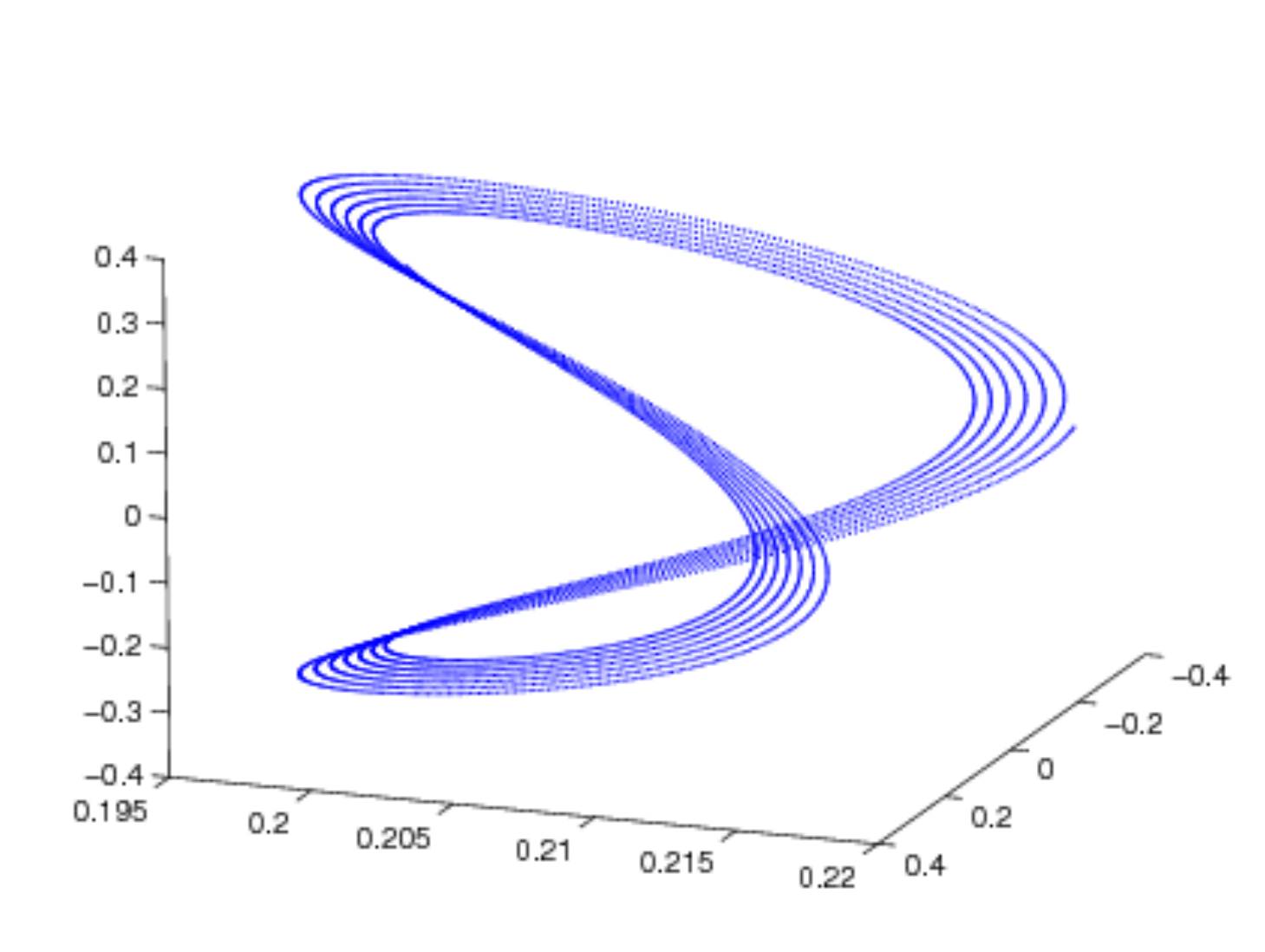}
\includegraphics[width=0.45\textwidth]{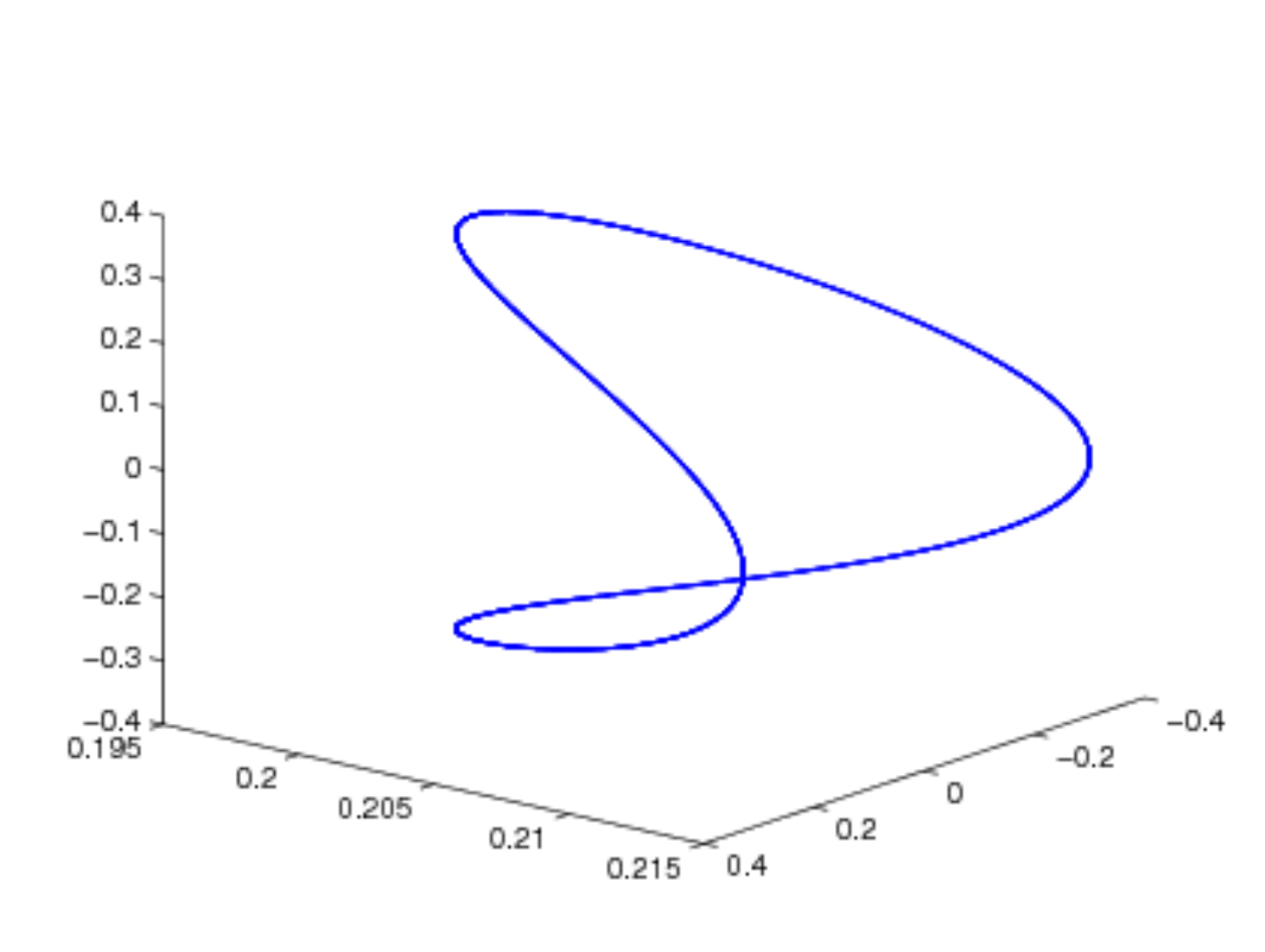}
\end{center}
\caption{Left: a spiralling orbit of the explicit Euler method
with the time-step $\epsilon=0.3$ applied to the Euler top. Right:
a single orbit of the Hirota-Kimura discretization with the same
time-step, lying on an invariant spatial elliptic curve
(intersection of two quadrics).}
\end{figure}

The Hirota-Kimura discretization of the Euler top
introduced in \cite{HK} reads as
\begin{equation}\label{eq: dET x}
\renewcommand{\arraystretch}{1.3}
\left\{\begin{array}{l}
\widetilde{x}_1-x_1=\epsilon\alpha_1(\widetilde{x}_2x_3+x_2\widetilde{x}_3),\\
\widetilde{x}_2-x_2=\epsilon\alpha_2(\widetilde{x}_3x_1+x_3\widetilde{x}_1),\\
\widetilde{x}_3-x_3=\epsilon\alpha_3(\widetilde{x}_1x_2+x_1\widetilde{x}_2).
\end{array}\right.
\end{equation}
Thus, the map $f:x\mapsto\widetilde{x}$ obtained by solving
(\ref{eq: dET x}) for $\widetilde{x}$, is given by:
\begin{equation}\label{eq: dET map}
\widetilde{x} =f(x,\epsilon)=A^{-1}(x,\epsilon)x, \qquad
A(x,\epsilon)=
\begin{pmatrix}
1 & -\epsilon\alpha_1 x_3 & -\epsilon\alpha_1 x_2 \\
-\epsilon\alpha_2 x_3 & 1 & -\epsilon\alpha_2 x_1 \\
-\epsilon\alpha_3 x_2 & -\epsilon\alpha_3 x_1 & 1
\end{pmatrix} .
\end{equation}
It might be instructive to have a look at the explicit formulas
for this map:
\begin{equation}
\label{eq: dET expl}
\left\{ \begin{array}{l}
\widetilde{x}_1=\dfrac{x_1+2\epsilon\alpha_1x_2x_3+
\epsilon^2x_1(-\alpha_2\alpha_3x_1^2+\alpha_3\alpha_1x_2^2+\alpha_1\alpha_2x_3^2)}
{\Delta(x,\epsilon)}\,,\\ \\
\widetilde{x}_2 = \dfrac{x_2+2\epsilon\alpha_2x_3x_1+
\epsilon^2x_2(\alpha_2\alpha_3x_1^2-\alpha_3\alpha_1x_2^2+\alpha_1\alpha_2x_3^2)}
{\Delta(x,\epsilon)}\,,\\ \\
\widetilde{x}_3 = \dfrac{x_3+2\epsilon\alpha_3x_1x_2+
\epsilon^2x_3(\alpha_2\alpha_3x_1^2+\alpha_3\alpha_1x_2^2-\alpha_1\alpha_2x_3^2)}
{\Delta(x,\epsilon)}\,,
\end{array}\right.
\end{equation}
where
\begin{equation}\nonumber
\Delta(x,\epsilon)=\det A(x,\epsilon)=
1-\epsilon^2(\alpha_2\alpha_3x_1^2+\alpha_3\alpha_1x_2^2+\alpha_1\alpha_2x_3^2)
-2\epsilon^3\alpha_1\alpha_2\alpha_3x_1x_2x_3.
\end{equation}
As always the case for HK-type discretizations, this map is
birational, and there holds the reversibility property:
\begin{equation}\nonumber
f^{-1}(x,\epsilon)=f(x,-\epsilon).
\end{equation}
Apart from the Lax representation which is still missing, the
discretization (\ref{eq: dET map}) exhibits all the usual features
of an integrable map: an invariant volume form, a bi-Hamiltonian
structure (that is, two compatible invariant Poisson structures),
two functionally independent conserved quantities in involution,
and solutions in terms of elliptic functions. The difference of
its qualitative behavior as compared with non-integrable
discretizations is striking, cf. Fig.~\ref{Fig: dET}. For further
details about the properties of this discretization we refer to
\cite{HK} and \cite{PS}.
The integrals have been first found in \cite{HK}, apparently with
the help of the approach discussed in the present work. However,
since the resulting integrals are sufficiently simple and nice,
their conservation can be easily verified by hands, therefore the
paper \cite{HK} presents them in an {\em ad hoc} form, without
explaining how they have been discovered. We now try to
reconstruct the way the results of \cite{HK} were originally
found. For this aim, we apply to the map (\ref{eq: dET map}) the
method described in section \ref{Sect: HK mechanism}.

\subsection{Hirota-Kimura bases}
Since all integrals of the Euler top are linear combinations of
the functions $x_k^2$, it is natural to try the set
\begin{equation}\label{eq: dET full basis}
\Phi=(x_1^2,x_2^2,x_3^2,1)
\end{equation}
as a HK-basis for the discrete time Euler top. An application of
the numerical algorithm (N) suggests that the following statement
holds:
\begin{theorem}\label{Th: dET full basis}
The set (\ref{eq: dET full basis}) is a HK-basis for the map
(\ref{eq: dET map}) with $\dim K_\Phi(x)=2$. Therefore, any orbit
of this map lies on the intersection of two quadrics in $\bbR^3$.
\end{theorem}
We will prove this theorem by finding two smaller HK-bases with
$d=1$. Namely, application of the numerical algorithm (N) suggests
that omitting any one of the four functions $1$, $x_k^2$ from the
basis $\Phi$ leads to a HK-basis with $d=1$. In other words, for
every $x\in\bbR^3$ there exists a one-dimensional space of vectors
$(c_1,c_2,c_3)$ such that
\begin{equation}\nonumber
(c_1x_1^2+c_2x_2^2+c_3x_3^2)\circ f^i(x)=0,\qquad i \in \bbZ,
\end{equation}
as well as a one-dimensional space of vectors $(d_1,d_2,d_4)$ such
that
\begin{equation}\nonumber
(d_1 x_1^2+d_2 x_2^2+d_4) \circ f^i(x)=0, \qquad i \in \bbZ.
\end{equation}
These numerical results can be now proven analytically.
\begin{prop}\label{Prop: dET first HK basis}
The set
\begin{equation}\nonumber
\Phi_0=(x_1^2,x_2^2,x_3^2)
\end{equation}
is a HK-basis for the map (\ref{eq: dET map}) with $\dim
K_{\Phi_0}(x)=1$. At each point $x\in\bbR^3$ there holds:
\[
K_{\Phi_0}(x)=[c_1:c_2:c_3]=[\, \alpha_3 x_2^2-\alpha_2 x_3^2 :
\alpha_1 x_3^2-\alpha_3 x_1^2 : \alpha_2 x_1^2- \alpha_1 x_2^2
\,].
\]
Setting $c_3=-1$, the functions
\begin{equation}\label{eq: dET c1 c2}
c_1(x)=\frac{\alpha_3 x_2^2-\alpha_2 x_3^2}{\alpha_1 x_2^2-\alpha_2 x_1^2},
\qquad
c_2(x)=\frac{\alpha_1 x_3^2-\alpha_3 x_1^2}{\alpha_1 x_2^2-\alpha_2 x_1^2}
\end{equation}
are integrals of motion of the map (\ref{eq: dET map}).
\end{prop}
{\bf Proof.} We proceed according to the recipe (B), set $c_3=-1$,
and solve symbolically the system
\begin{equation}
\label{eq: dET_ansatz_0}
(c_1x_1^2+c_2x_2^2)\circ f^i(x)=x_3^2\circ f^i(x), \quad i=0,1,
\end{equation}
which involves two non-homogeneous equations for two unknowns.
System (\ref{eq: dET_ansatz_0}) can be written as
\begin{equation}\label{eq: dET_ansatz_0 long}
\renewcommand{\arraystretch}{1.3}
\left\{ \begin{array}{ccl}
c_1x_1^2+c_2x_2^2 & = & x_3^2,\\
c_1\widetilde{x}_1^2+c_2\widetilde{x}_2^2 & = &
\widetilde{x}_3^2,\end{array}\right.
\end{equation}
where, of course, explicit formulas (\ref{eq: dET expl}) have to
be used for $\widetilde{x}_k$. The solution of this system is
given by formulas (\ref{eq: dET c1 c2}). The components of the
solution do not depend on $\epsilon$, therefore, according to the
recipe (D), we conclude that functions (\ref{eq: dET c1 c2}) are
integrals of motion of the map (\ref{eq: dET map}). \hfill $\Box$

It should be mentioned that the independence of the solution
$(c_1,c_2)$ on $\epsilon$, or, more generally, the dependence
through even powers of $\epsilon$ only, which will be mentioned on
many occasions below, starting with Proposition \ref{Prop: dET
second HK basis}, is not granted by any well-understood mechanism.
Rather, it is just an instance of very {\em remarkable and
miraculous cancellations} of non-even polynomials. We illustrate
this phenomenon by providing additional details to the previous
proof. The solution of eqs. \eqref{eq: dET_ansatz_0 long} by the
Cramer's rule is given by ratios of determinants of the type
\begin{equation}\label{eq: dET for cancel}
\renewcommand{\arraystretch}{1.3}
\left|\begin{array}{cc} x_i^2 & x_j^2 \\ \widetilde{x}_i^2 &
\widetilde{x}_j^2\end{array}\right|=\frac{4\epsilon(\alpha_jx_i^2-\alpha_ix_j^2)
(x_1+\epsilon\alpha_1x_2x_3)(x_2+\epsilon\alpha_2x_3x_1)(x_3+\epsilon\alpha_3x_1x_2)}
{\Delta^2(x,\epsilon)}
\end{equation}
In the ratios of such determinants everything cancels out, except
for the factors $\alpha_jx_i^2-\alpha_ix_j^2$. The cancellation of
the denominators $\Delta^2(x,\epsilon)$ is, of course, no wonder,
but the cancellation of the non-even factors in the numerators is
rather miraculous.

One more typical phenomenon occurs in Proposition \ref{Prop: dET
first HK basis}: although we have found apparently two integrals
of motion (\ref{eq: dET c1 c2}), they turn out to be {\em
functionally dependent}. Indeed, there holds an identity
\[
\alpha_1 c_1(x)+\alpha_2 c_2(x)=\alpha_3,
\]
so that for each $x\in\bbR^3$ the space $K_{\Phi_0}(x)$ is
orthogonal to the constant vector $(\alpha_1,\alpha_2,\alpha_3)$.
If one would have guessed this relation numerically, one could
simplify the computation of the integrals $c_1,c_2$ by considering
the system
\begin{equation}\label{eq: dET_ansatz_0 short}
\left\{ \begin{array}{ccl} c_1x_1^2+c_2x_2^2 & = & x_3^2,\\
c_1\alpha_1+c_2\alpha_2 & = & \alpha_3,\end{array}\right.
\end{equation}
instead of (\ref{eq: dET_ansatz_0 long}). Observe that existence
of a linear relation allows one to reduce a number of iterates of
$f$ involved in the linear system (in the present situation, the
system (\ref{eq: dET_ansatz_0 short}) contains no iterates of $f$
at all!). The latter system would lead to the same formulas
(\ref{eq: dET c1 c2}), however, in this case one could not argue
as in (D), and would be forced to prove that the functions
(\ref{eq: dET c1 c2}) are integrals of motion directly, by
verifying for them equations (\ref{eq: integrals}).

Anyway, the existence of the HK-basis $\Phi_0$ yields existence of
only one independent integral of the map $f$, which is not enough
to assure the integrability of $f$.

\begin{prop}\label{Prop: dET second HK basis}
The set
\begin{equation}\nonumber
\Phi_1=(x_1^2,x_2^2,1)
\end{equation}
is a HK-basis for the map (\ref{eq: dET map}) with $\dim
K_{\Phi_1}(x)=1$. At each point $x\in\bbR^3$ there holds:
\[
K_{\Phi_1}(x)=[d_1 : d_2 : -1],
\]
where
\begin{equation}\label{eq: dET c4 c5}
d_1(x)=\frac{\alpha_2(1-\epsilon^2\alpha_3\alpha_1 x_2^2)}
{\alpha_2 x_1^2-\alpha_1 x_2^2}\,, \qquad
d_2(x)=\frac{\alpha_1(1-\epsilon^2\alpha_2\alpha_3 x_1^2)}
{\alpha_1 x_2^2-\alpha_2 x_1^2}\,.
\end{equation}
These functions are integrals of motion of the map (\ref{eq: dET
map}).
\end{prop}
{\bf Proof.} Following again prescription (B), we set $d_4=-1$,
and solve symbolically the non-homogeneous system
\begin{equation}\nonumber
(d_1 x_1^2+d_2 x_2^2) \circ f^i(x)=1, \quad i=0,1,
\end{equation}
or
\[
\renewcommand{\arraystretch}{1.3}
\left\{ \begin{array}{ccl} d_1 x_1^2+d_2 x_2^2 & = & 1,\\
d_1\widetilde{x}_1^2+d_2\widetilde{x}_2^2 & = &
1.\end{array}\right.
\]
The solution is given by eq. (\ref{eq: dET c4 c5}), due to eq.
\eqref{eq: dET for cancel} and
\[
\renewcommand{\arraystretch}{1.3}
\left|\begin{array}{cc} 1 & x_i^2 \\ 1 &
\widetilde{x}_i^2\end{array}\right|
=\frac{4\epsilon\alpha_i(1-\epsilon^2\alpha_j\alpha_kx_i^2)
(x_1+\epsilon\alpha_1x_2x_3)(x_2+\epsilon\alpha_2x_3x_1)(x_3+\epsilon\alpha_3x_1x_2)}
{\Delta^2(x,\epsilon)}
\]
This time its components do depend on $\epsilon$, but are
manifestly even functions of $\epsilon$. Everything non-even
luckily cancels, again.  Therefore, the argument (D) is still
applicable, so that the functions (\ref{eq: dET c4 c5}) are
integrals of motion of the map $f$. \hfill $\Box$

Functions (\ref{eq: dET c4 c5}) are again functionally dependent,
because of
\[
\alpha_1d_1(x)+\alpha_2d_2(x)=\epsilon^2\alpha_1\alpha_2\alpha_3.
\]
However, they are, clearly, functionally independent on the
previously found functions (\ref{eq: dET c1 c2}), because
$c_1,c_2$ depend on $x_3$, while $d_1,d_2$ do not.
\smallskip

Of course, the permutational symmetry yields that each of the sets
of monomials $\Phi_2=(x_2^2,x_3^2,1)$ and $\Phi_3=(x_1^2,x_3^2,1)$
is a HK-basis, as well, with $\dim K_{\Phi_2}(x)=\dim
K_{\Phi_3}(x)=1$. Any two of the four found one-dimensional
null-spaces span the full null-space $K_\Phi(x)$. In particular,
$K_{\Phi_0}(x)$ lies in $K_{\Phi_1}(x)\oplus K_{\Phi_2}(x)$.
\medskip

Summarizing, we have found a HK-basis with a two-dimensional
null-space, as well as two functionally independent conserved
quantities for the HK-discretization of the Euler top. Both
results yield integrability of this discretization, in the sense
that its orbits are confined to closed curves in $\bbR^3$.
Moreover, each such curve is an intersection of two quadrics,
which in the general position case is an elliptic curve.

\section{Hirota-Kimura-type discretization of the Clebsch system}
\label{Sect: Clebsch1}

\subsection{Clebsch system}
The motion of a rigid body in an ideal fluid can be described by
the so called {\em Kirchhoff equations} \cite{Kirch}:
\begin{equation}\label{eq: Kirch}
\renewcommand{\arraystretch}{2.2}
\left\{\begin{array}{l} \dot{m}=m\times\dfrac{\partial H}{\partial
m}
        +p\times\dfrac{\partial H}{\partial p}, \\
\dot{p}=p\times\dfrac{\partial H}{\partial m},
\end{array}\right.
\end{equation}
with $H$ being a quadratic form in $m=(m_1,m_2,m_3)\in\bbR^3$ and
$p=(p_1,p_2,p_3)\in\bbR^3$; here $\times$ denotes vector product
in $\bbR^3$. The physical meaning of $m$ is the total angular
momentum, whereas $p$ represents the total linear momentum of the
system. System (\ref{eq: Kirch}) is Hamiltonian with the Hamilton
function $H(m,p)$, with respect to the Poisson bracket
\begin{equation}\label{eq: e3 bracket}
\{m_i,m_j\}=m_k,\qquad \{m_i,p_j\}=p_k,
\end{equation}
where $(i,j,k)$ is a cyclic permutation of (1,2,3) (all other
pairwise Poisson brackets of the coordinate functions are obtained
from these by the skew-symmetry, or otherwise vanish). A detailed
introduction to the general context of rigid body dynamics and its
mathematical foundations can be found in \cite{MR}.

A famous integrable case of the Kirchhoff equations was discovered
in \cite{C} and is characterized by the Hamilton function
$H=\frac{1}{2}\sum_{i=1}^{3}(m_i^2+\omega_ip_i^2)$. The
corresponding equations of motion read:
\begin{equation}\nonumber
\left\{\begin{array}{l}
\dot{m}=p\times\Omega p,\\
\dot{p}=p\times m,
\end{array}\right.
\end{equation}
where $\Omega={\rm diag}(\omega_1,\omega_2,\omega_3)$ is the
matrix of parameters, or in components:
\begin{eqnarray}
\dot{m}_1 & = & (\omega_3-\omega_2)p_2p_3, \nonumber  \\
\dot{m}_2 & = & (\omega_1-\omega_3)p_3p_1, \nonumber \\
\dot{m}_3 & = & (\omega_2-\omega_1)p_1p_2, \nonumber \\
\dot{p}_1 & = & m_3 p_2 - m_2 p_3,  \nonumber\\
\dot{p}_2 & = & m_1 p_3 - m_3 p_1,  \nonumber \\
\dot{p}_3 & = & m_2 p_1 - m_1 p_2.  \nonumber
\end{eqnarray}
This is the system which will be called the \textit{Clebsch
system} hereafter. For an embedding of this system into the modern
theory of integrable systems see \cite{Per, RSTS}. The Clebsch
system possesses four independent quadratic integrals:
\begin{eqnarray}
H_1 & = & m_1^2+m_2^2 +m_3^2+\omega_1p_1^2+\omega_2p_2^2+\omega_3p_3^2,
                                   \label{eq: Clebsch H1}
\\
H_2 & = & \omega_1m_1^2+\omega_2m_2^2+\omega_3m_3^2 -
         \omega_2\omega_3p_1^2-\omega_3\omega_1p_2^2-\omega_1\omega_2p_3^2,
                                   \label{eq: Clebsch H2}
\\
H_3 & = & p_1^2+p_2^2+p_3^2,
\label{eq: Clebsch C1}
\\
H_4 & = & m_1p_1+m_2p_2+m_3p_3.
\label{eq: Clebsch C2}
\end{eqnarray}
These integrals are in involution with respect to the bracket
(\ref{eq: e3 bracket}), moreover, $H_3,H_4$ are its Casimir
functions (are in involution with any function on the phase
space). However, the Hamiltonian structure will not play any role
in the present paper. The set of linear combinations of the
quadratic Hamiltonians $H_1,H_2,H_3$ coincides with the set of
linear combinations of the functions
\bea
I_1&=&p_1^2+\frac{m_2^2}{\omega_1-\omega_3}+\frac{m_3^2}{\omega_1-\omega_2}\,,\nonumber \\
I_2&=&p_2^2+\frac{m_1^2}{\omega_2-\omega_3}+\frac{m_3^2}{\omega_2-\omega_1}\,,\nonumber \\
\label{eq: Clebsch I}
I_3&=&p_3^2+\frac{m_1^2}{\omega_3-\omega_2}+\frac{m_2^2}{\omega_3-\omega_1}\,. \nonumber
\eea
For instance,
\[
H_1=\omega_1I_1+\omega_2I_2+\omega_3I_3,\quad
H_1=-\omega_2\omega_3I_1-\omega_3\omega_1I_2-\omega_1\omega_2I_3,\quad
H_3=I_1+I_2+I_3.
\]

\subsection{Discrete equations of motion}
Applying the Hirota-Kimura (or Kahan) approach to the Clebsch
system, we arrive at the following discretization, proposed in
\cite{RC}:
\begin{eqnarray}\label{eq: dC}
\widetilde{m}_1-m_1 & = & \epsilon(\omega_3-\omega_2)
(\widetilde{p}_2p_3+p_2\widetilde{p}_3),         \nonumber \\
\widetilde{m}_2-m_2 & = & \epsilon(\omega_1-\omega_3)
(\widetilde{p}_3p_1+p_3\widetilde{p}_1),         \nonumber \\
\widetilde{m}_3-m_3 & = & \epsilon(\omega_2-\omega_1)
(\widetilde{p}_1p_2+p_1\widetilde{p}_2),         \nonumber \\
\widetilde{p}_1-p_1 & = &
\epsilon(\widetilde{m}_3p_2+m_3\widetilde{p}_2)-
\epsilon(\widetilde{m}_2p_3+m_2\widetilde{p}_3), \nonumber \\
\widetilde{p}_2-p_2 & = &
\epsilon(\widetilde{m}_1p_3+m_1\widetilde{p}_3)-
\epsilon(\widetilde{m}_3p_1+m_3\widetilde{p}_1), \nonumber \\
\widetilde{p}_3-p_3 & = &
\epsilon(\widetilde{m}_2p_1+m_2\widetilde{p}_1)-
\epsilon(\widetilde{m}_1p_2+m_1\widetilde{p}_2).\nonumber
\end{eqnarray}
In matrix form this can be put as
\[
M(m,p,\epsilon)
\begin{pmatrix} \widetilde{m} \\ \widetilde{p}\end{pmatrix} =
\begin{pmatrix} m \\ p \end{pmatrix} ,
\]
where
\[
M(m,p,\epsilon) = \begin{pmatrix}
1 & 0 & 0 & 0 & \epsilon\omega_{23}p_3 & \epsilon\omega_{23}p_2 \\
0 & 1 & 0 & \epsilon\omega_{31}p_3 & 0 & \epsilon\omega_{31}p_1 \\
0 & 0 & 1 & \epsilon\omega_{12}p_2 & \epsilon\omega_{12}p_1 & 0 \\
0 & \epsilon p_3 & -\epsilon p_2 & 1 & -\epsilon m_3 & \epsilon m_2 \\
-\epsilon p_3 & 0 & \epsilon p_1 & \epsilon m_3 & 1 & -\epsilon m_1  \\
\epsilon p_2 & -\epsilon p_1 & 0 & -\epsilon m_2 & \epsilon m_1 & 1
\end{pmatrix},
\]
and the abbreviation $\omega_{ij}=\omega_i-\omega_j$ is used. The
solution of this $6\times 6$ linear system yields the birational
map $f:\bbR^6\to\bbR^6$,
\begin{equation}\label{eq: dC map}
\begin{pmatrix} \widetilde{m} \\ \widetilde{p}\end{pmatrix}=f(m,p,\epsilon)
= M^{-1}(m,p,\epsilon)\begin{pmatrix} m \\ p \end{pmatrix},
\end{equation}
called hereafter the \textit{discrete Clebsch system}. As usual,
the reversibility property holds:
\begin{equation}
\label{eq: dC reverse}
f^{-1}(m,p,\epsilon) = f(m,p,-\epsilon).
\end{equation}

A remark on the complexity of the iterates of $f$ is in order
here. Each component of $(\widetilde m,\widetilde p)=f(m,p)$ is a
rational function with the numerator and the denominator being
polynomials on $m_k,p_k$ of total degree 6. The numerators of
$\widetilde{p}_k$ consist of 31 monomials, the numerators of
$\widetilde{m}_k$ consist of 41 monomials, the common denominator
consists of 28 monomials. It should be taken into account that the
coefficients of all these polynomials depend, in turn,
polynomially on $\epsilon$ and $\omega_k$, which additionally
increases their complexity for a symbolic manipulator. Expressions
for the second iterate swell to astronomical length prohibiting
naive attempts to compute them symbolically. Using MAPLE's
LargeExpressions package \cite{LE} and an appropriate veiling
strategy it is however possible to obtain $f^2(m, p)$ with a
reasonable amount of memory. Some impression on the complexity can
be obtained from Table \ref{Table f2}. The resulting expressions
are too big to be used in further symbolic computations. Consider,
for instance, the numerator of the $p_1$-component of $f^2(m,p)$.
As a polynomial of $m_k,p_k$, it contains 64 056 monomials; their
coefficients are, in turn, polynomials of $\epsilon$ and
$\omega_k$, and, considered as a polynomial of the phase variables
and the parameters, this expression contains 1 647 595 terms.

\begin{table}[htbp]
   \begin{tabular}{|c||ccccccc|}
     \hline
     & deg  & $\deg_{p_1}$ & $\deg_{p_2}$ & $\deg_{p_3}$ &
                       $\deg_{m_1}$ & $\deg_{m_2}$ & $\deg_{m_3}$  \\

     \hline\hline
      Common denominator of $f^2$       & 27 & 24 & 24 & 24 & 12 & 12 & 12   \\
      Numerator of $p_1$-comp. of $f^2$ & 27 & 25 & 24 & 24 & 12 & 12 & 12  \\
      Numerator of $p_2$-comp. of $f^2$ & 27 & 24 & 25 & 24 & 12 & 12 & 12  \\
      Numerator of $p_3$-comp. of $f^2$ & 27 & 24 & 24 & 25 & 12 & 12 & 12  \\
      Numerator of $m_1$-comp. of $f^2$ & 33 & 28 & 28 & 28 & 15 & 14 & 14  \\
      Numerator of $m_2$-comp. of $f^2$ & 33 & 28 & 28 & 28 & 14 & 15 & 14  \\
      Numerator of $m_3$-comp. of $f^2$ & 33 & 28 & 28 & 28 & 14 & 14 & 15  \\
      \hline
   \end{tabular}
   \vspace{1mm}
   \caption{Degrees of the numerators and the denominator of the second iterate
   $f^2(m,p)$}
\label{Table f2}
\end{table}

\subsection{Phase portrait and integrability}

\begin{figure}[hbt]
\centering \subfigure[$m_1,m_2,m_3$] { \scalebox{0.4} {
\includegraphics{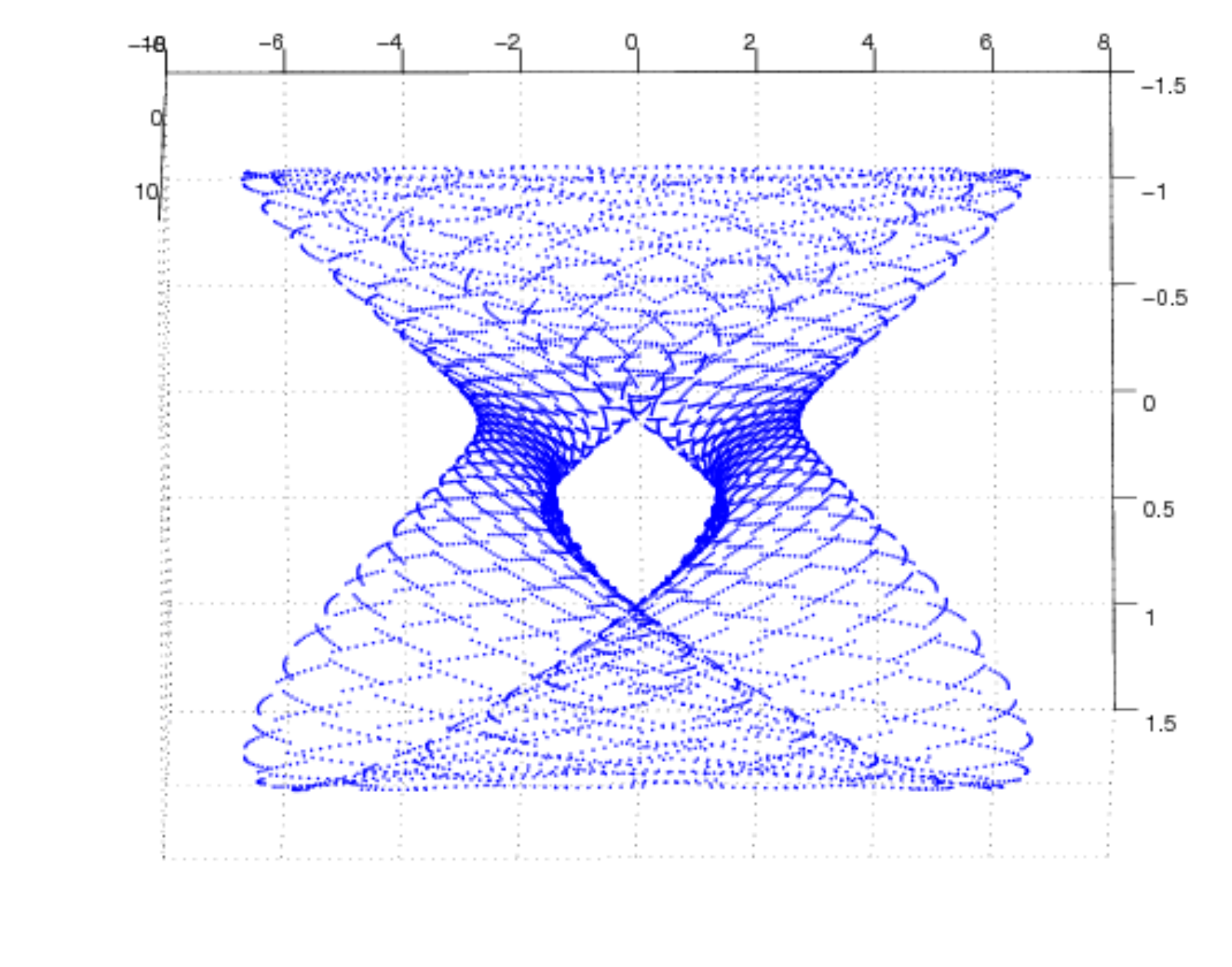} } }
\subfigure[$p_1,p_2,p_3$] { \scalebox{0.4} {
\includegraphics{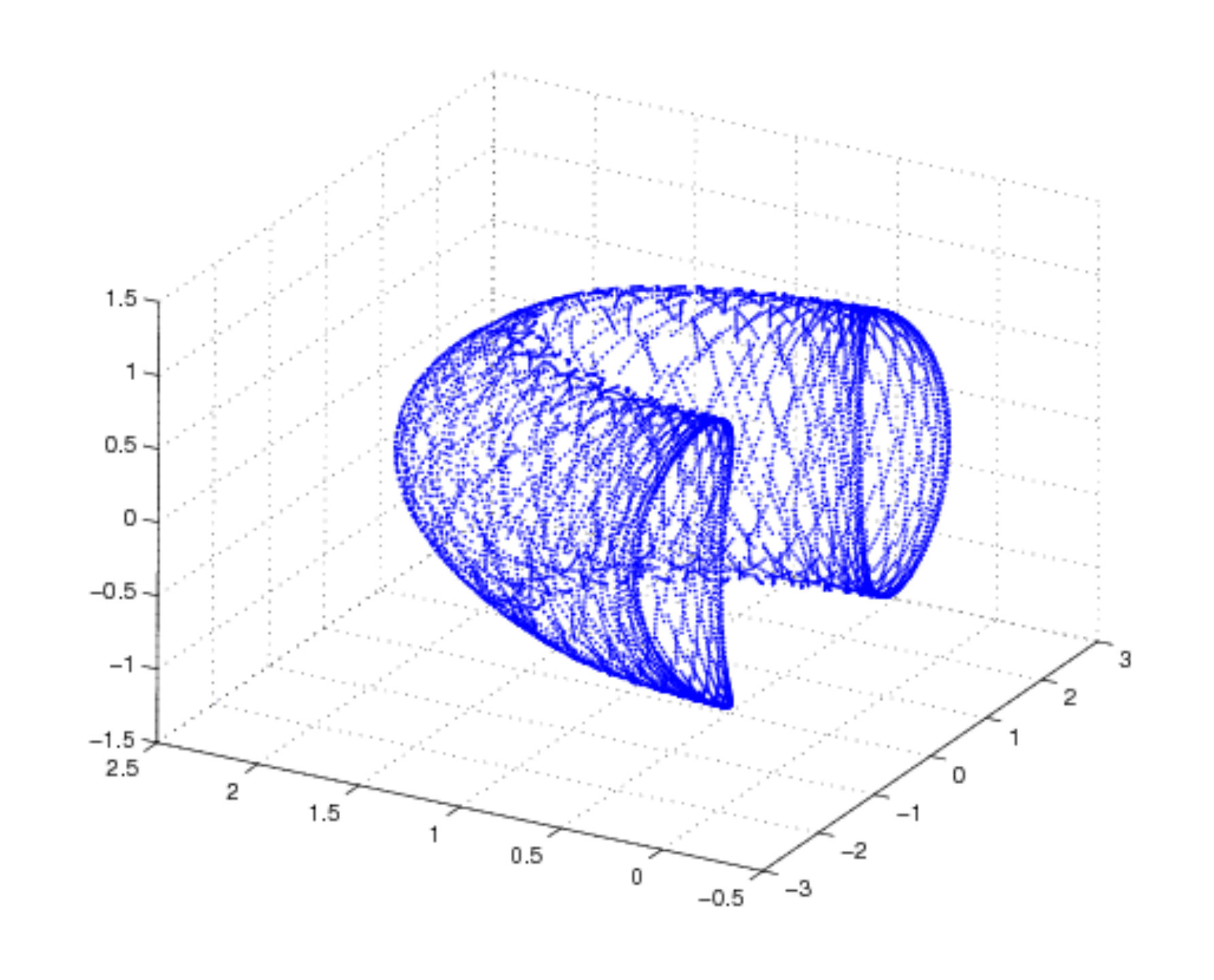} } }
\caption{An orbit of the discrete Clebsch system with
$\omega_1=1$, $\omega_2=0.2$, $\omega_3=30$ and $\epsilon=1$;
initial point $(m_0,p_0)=(1,1,1,1,1,1)$.} \label{plot1}
\end{figure}

\begin{figure}[hbt]
\centering \subfigure[$m_1,m_2,m_3$] { \scalebox{0.38} {
\includegraphics{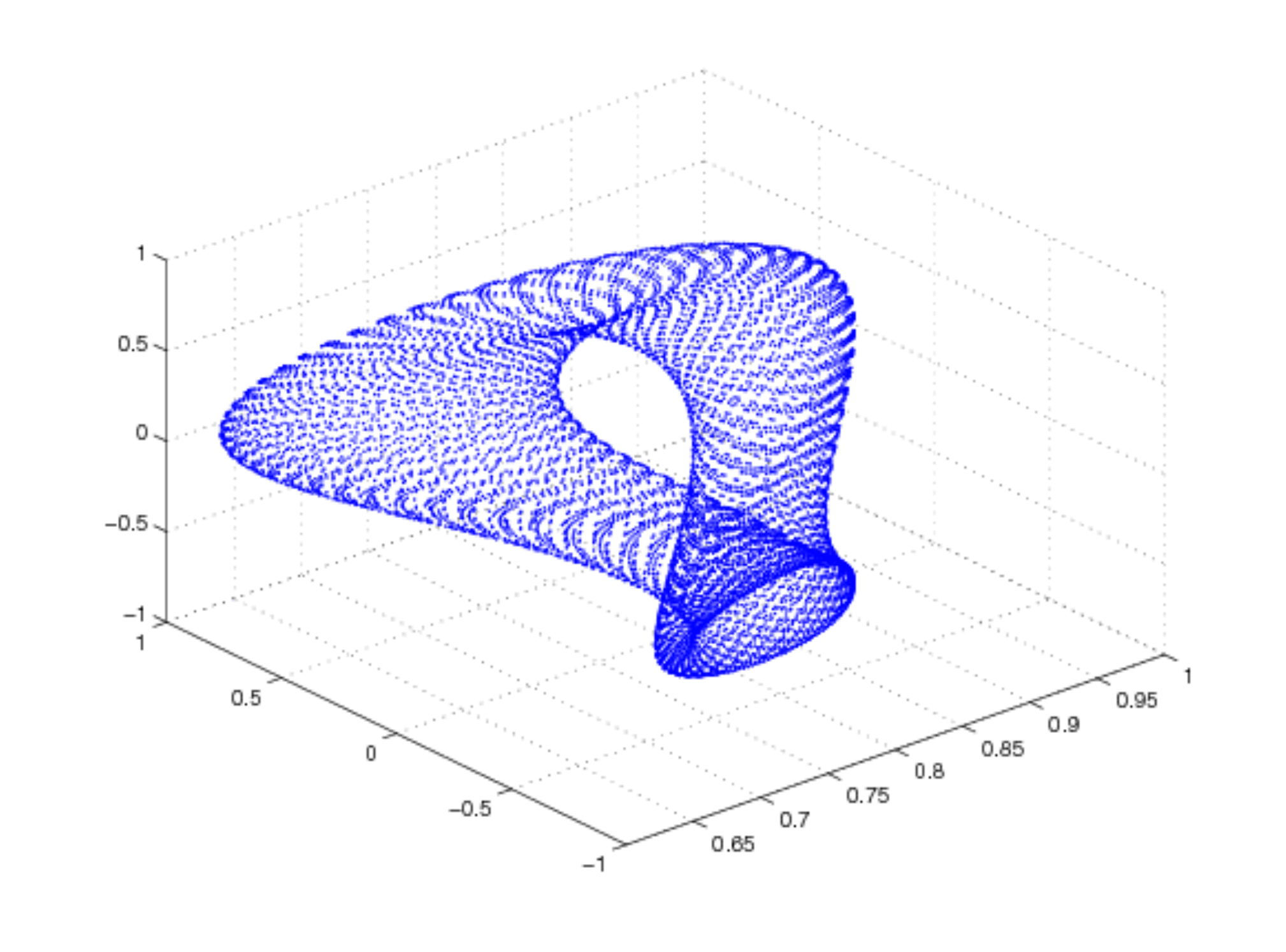} } }
\subfigure[$p_1,p_2,p_3$] { \scalebox{0.4} {
\includegraphics{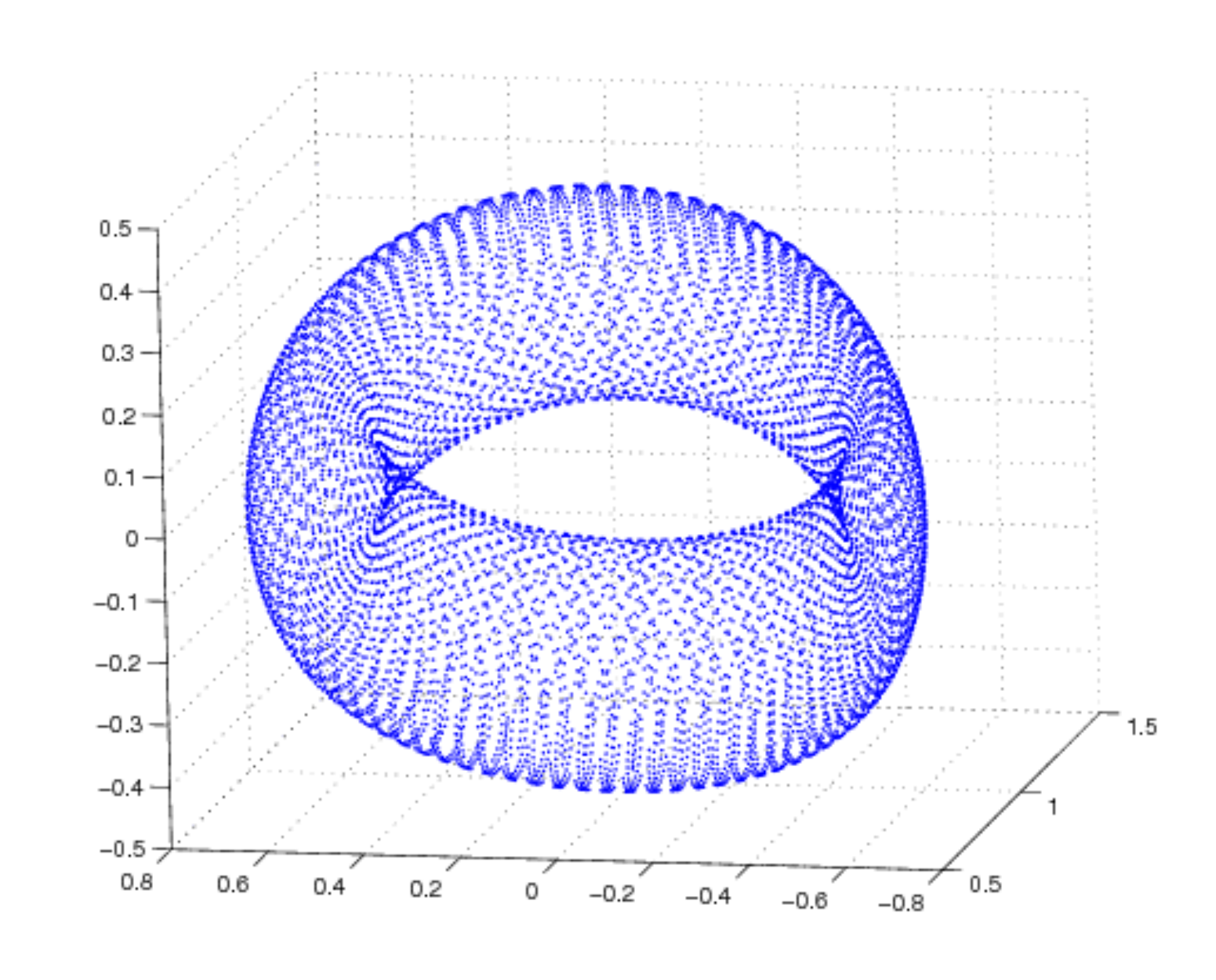} } }
\caption{An orbit of the discrete Clebsch system with
$\omega_1=0.1$, $\omega_2=0.2$, $\omega_3=0.3$ and $\epsilon=1$;
initial point $(m_0,p_0)=(1,1,1,1,1,1)$.} \label{plot2}
\end{figure}

We now address the problem whether the discrete Clebsch system is
integrable. Figs. \ref{plot1} and \ref{plot2} show plots of the
discrete Clebsch system (\ref{eq: dC map}), produced with MATLAB,
for two different sets of parameters values. These plots indicate
a quite regular behavior of the orbits of the discrete Clebsch
system. Each orbit seems to fill out a two-dimensional surface in
the 6-dimensional phase space. Leaving aside the Hamiltonian
aspects of integrability, we are interested just in this simpler
issue: do orbits of the map (\ref{eq: dC map}) lie on
two-dimensional surfaces in $\bbR^6$? A usual way to establish
such a property would be to establish the existence of four
functionally independent conserved quantities for this map. (We
note in passing that plots of orbits are not very reliable in
deciding about integrability. For instance, there are indications
that the Kahan's discretization (\ref{eq: Kahan LV}) of
the Lotka-Volterra system is non-integrable, even if its orbits
visually lie on closed curves in the phase plane. A strong
magnification unveils the existence of very small regions in the
phase plane with a chaotic behavior.)


We will show that the answer to the above question is in
affirmative. For this aim, we apply the approach based on the
notion of HK-basis. As a first step, we apply the numerical
algorithm (N) to the {\em maximal set of monomials}, which
includes all monomials of which the integrals (\ref{eq: Clebsch
H1})--(\ref{eq: Clebsch C2}) of the continuous Clebsch system are
built:
\begin{eqnarray*}
&\varphi_1(m,p)=p_1^2,\quad \varphi_2(m,p)=p_2^2, \quad \varphi_3(m,p)=p_3^2,&\\
&\varphi_4(m,p)=m_1^2,\quad \varphi_5(m,p)=m_2^2, \quad \varphi_6(m,p)=m_3^2,&\\
&\varphi_7(m,p)=m_1p_1,\quad \varphi_8(m,p)=m_2p_2,\quad \varphi_9(m,p)=m_3p_3,&\\
& \varphi_{10}(m,p)=1. &
\end{eqnarray*}
We come to the following result:
\begin{theorem}\label{Th: dClebsch max basis}
The set of functions
\begin{equation}\nonumber
\Phi=(p_1^2,p_2^2,p_3^2,m_1^2,m_2^2,m_3^2,m_1p_1,m_2p_2,m_3p_3,1)
\end{equation}
is a HK-basis for the map (\ref{eq: dC map}), with $\dim
K_\Phi(m,p)=4$. Thus, any orbit of the map (\ref{eq: dC map}) lies
on an intersection of four quadrics in $\bbR^6$.
\end{theorem}
At this point Theorem \ref{Th: dClebsch max basis} remains a
numerical result, based on the algorithm (N). A direct symbolical
proof of this statement is impossible, since it requires dealing
with $f^i$, $i\in[-4,4]$, and the fourth iterate $f^4$ is a
forbiddingly large expression. In order to {\em prove} Theorem
\ref{Th: dClebsch max basis} and to extract from it four
independent integrals of motion, it is desirable to find
HK-(sub)bases with a smaller number of monomials, corresponding to
some (preferably one-dimensional) subspaces of $K_\Phi(m,p)$. A
much more detailed information on the HK-bases is provided by the
following statement.
\begin{theorem}\label{Th: dClebsch 1dim bases}
The following four sets of functions are HK-bases for the map
(\ref{eq: dC map}) with one-dimensional null-spaces:
\begin{eqnarray}
\Phi_0 & = & (p_1^2,p_2^2,p_3^2,1),
\label{eq: dC basis 1}\\
\Phi_1 & = & (p_1^2,p_2^2,p_3^2,m_1^2,m_2^2,m_3^2,m_1p_1),
\label{eq: dC basis Psi1}\\
\Phi_2 & = & (p_1^2,p_2^2,p_3^2,m_1^2,m_2^2,m_3^2,m_2p_2),
\label{eq: dC basis Psi2}\\
\Phi_3 & = & (p_1^2,p_2^2,p_3^2,m_1^2,m_2^2,m_3^2,m_3p_3).
\label{eq: dC basis Psi3}
\end{eqnarray}
If all the null-spaces are considered as subspaces of
$\,\bbR^{10}$, so that
\begin{equation*}
\begin{array}{cclrrrrrrrrr}
K_{\Phi_0} & = & [c_1: & c_2: & c_3: & 0: & 0: & 0: & 0: & 0: & 0:
& c_{10}],\\
K_{\Phi_1} & = & [\alpha_1: & \alpha_2: & \alpha_3: & \alpha_4: &
\alpha_5: & \alpha_6: & \alpha_7: & 0: & 0: & 0],\\
K_{\Phi_2} & = & [\beta_1: & \beta_2: & \beta_3: & \beta_4: &
\beta_5: & \beta_6: & 0: & \beta_8: & 0: & 0],\\
K_{\Phi_3} & = & [\gamma_1: & \gamma_2: & \gamma_3: & \gamma_4: &
\gamma_5: & \gamma_6: & 0: & 0: & \gamma_9: & 0],
\end{array}
\end{equation*}
then there holds:
\begin{equation} \nonumber
K_{\Phi}=K_{\Phi_0}\oplus K_{\Phi_1}\oplus K_{\Phi_2}\oplus
K_{\Phi_3}.
\end{equation}
\end{theorem}
Also this statement was first found with the help of numerical
experiments based on the algorithm (N). In what follows, we will
discuss how these claims can be given a rigorous (computer
assisted) proof, and how much additional information (for
instance, about conserved quantities for the map (\ref{eq: dC
map})) can be extracted from such a proof.

\subsection{First HK-basis}

\begin{theorem}\label{Th: dC basis 1}
The set (\ref{eq: dC basis 1}) is a HK-basis for the map (\ref{eq:
dC map}) with $\dim K_{\Phi_0}(m,p)=1$. At each point
$(m,p)\in\bbR^6$ there holds:
\begin{eqnarray}
\lefteqn{K_{\Phi_0}(m,p)=[c_1:c_2:c_3:c_{10}]}\nonumber\\
&=&\left[\,
\dfrac{1+\epsilon^2(\omega_1-\omega_2)p_2^2+\epsilon^2(\omega_1-\omega_3)p_3^2}
{p_1^2+p_2^2+p_3^2}:
\dfrac{1+\epsilon^2(\omega_2-\omega_1)p_1^2+\epsilon^2(\omega_2-\omega_3)p_3^2}
{p_1^2+p_2^2+p_3^2}: \right.
\nonumber\\
&& \left.
\dfrac{1+\epsilon^2(\omega_3-\omega_1)p_1^2+\epsilon^2(\omega_3-\omega_2)p_2^2}
{p_1^2+p_2^2+p_3^2}: -1\,\right]
\nonumber
\\
& = &
\left[\,\frac{1}{J}+\epsilon^2\omega_1:\frac{1}{J}+\epsilon^2\omega_2:
\frac{1}{J}+\epsilon^2\omega_3:-1\,\right],\label{eq: dC ans1 res}
\end{eqnarray}
where
\begin{equation}
\label{eq: dC C1} J(m,p,\epsilon)=\frac{p_1^2+p_2^2+p_3^2}
{1-\epsilon^2(\omega_1p_1^2+\omega_2p_2^2+\omega_3p_3^2)}\,.
\end{equation}
The function (\ref{eq: dC C1}) is an integral of motion of the map
(\ref{eq: dC map}).
\end{theorem}
{\bf Proof.} The statement of the theorem means that for every
$(m,p)\in\bbR^6$ the space of solutions of the homogeneous system
\begin{equation}\nonumber
(c_1p_1^2+c_2p_2^2+c_3p_3^2+c_{10})\circ f^i(m,p)=0, \qquad
i=0,\ldots,3,
\end{equation}
is one-dimensional. This system involves the third iterate of $f$,
therefore its symbolical treatment is impossible. According to the
strategy (B), we set $c_{10}=-1$ and consider the non-homogeneous
system
\begin{equation}\label{eq: clebsch_ansatz_1 nonhom}
(c_1p_1^2+c_2p_2^2+c_3p_3^2)\circ f^i(m,p)=1, \qquad i=0,1,2.
\end{equation}
This system involves the second iterate of $f$, which still
precludes its symbolical treatment. There are now several
possibilities to proceed.

\begin{itemize}
\item First, we could follow the recipe (E) and find further
information about the solutions $c_i$. For this aim, we plot the
points $(c_1(m,p),c_2(m,p),c_3(m,p))$ for different initial data
$(m,p)\in\bbR^6$. Figure \ref{plot3} shows such a plot, with 300
initial data $(m,p)$ randomly chosen from the set $[0,1]^6$.
\begin{figure}[hbt]
\centering \scalebox{0.6} {
\includegraphics{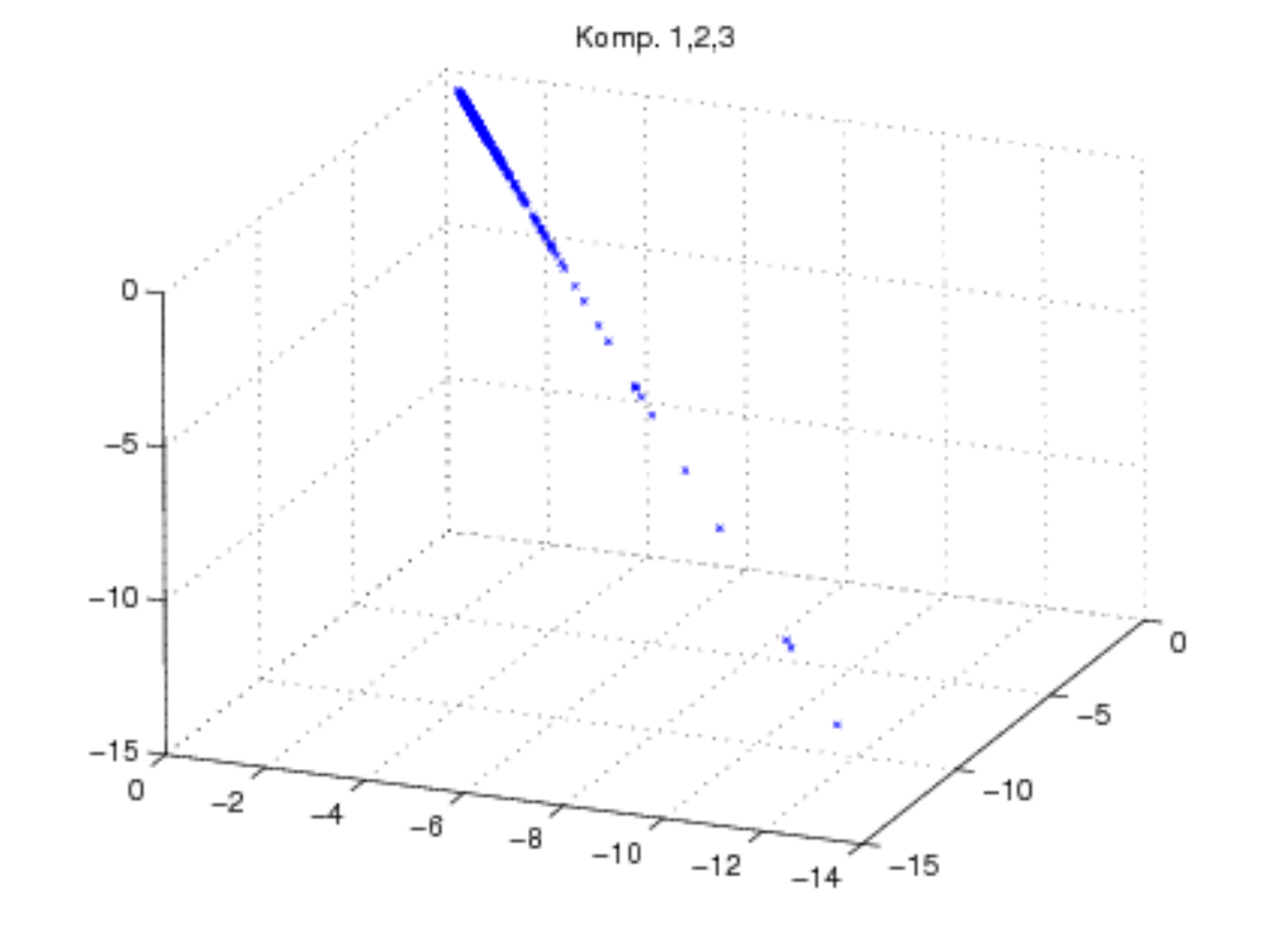} }
\caption{Plot of the coefficients $c_1,c_2,c_3$} \label{plot3}
\end{figure}
The points $(c_1(m,p),c_2(m,p),c_3(m,p))$ seem to lie on a line in
$\bbR^3$, which means that there should be two linear dependencies
between the functions $c_1,c_2$ and $c_3$. In order to identify
these linear dependencies, we run the PSLQ algorithm \cite{PSLQ,
CPSLQ} with the vectors $(c_1,c_2,1)$ as input (see Remark after
the end of the proof, concerning implementation of this step). On
this way we obtain the conjecture
\[
c_1-c_2=\epsilon^2(\omega_1-\omega_2).
\]
Similarly, running the PSLQ algorithm with the vectors
$(c_2,c_3,1)$ as input leads to the conjecture
\[
c_2-c_3=\epsilon^2(\omega_2-\omega_3).
\]
Having identified (numerically!) these two linear relations, we
use them instead of two equations in the system (\ref{eq:
clebsch_ansatz_1 nonhom}), say the equations for $i=1,2$. The
resulting system becomes extremely simple:
\[
\left\{\begin{array}{l} c_1p_1^2+c_2p_2^2+c_3p_3^2=1,\\
c_1-c_2=\epsilon^2(\omega_1-\omega_2),\\
c_2-c_3=\epsilon^2(\omega_2-\omega_3).
\end{array}\right.
\]
It contains no iterates of $f$ at all and can be solved
immediately by hands, with the result (\ref{eq: dC ans1 res}). It
should be stressed that this result still remains conjectural, and
one has to prove {\em a posteriori} that the functions
$c_1,c_2,c_3$ are integrals of motion.

\item Alternatively, we can combine the above approach based on
the prescription (E) with the recipe (D). For this, we use just
one of the linear dependencies found above to replace the equation
in (\ref{eq: clebsch_ansatz_1 nonhom}) with $i=2$, and then let
MAPLE solve the remaining system. The computation takes 22,33
secs. on a 1.83 Ghz Core Duo PC  and consumes 32,43 MB RAM. The
output is still as in (\ref{eq: dC ans1 res}), but arguing this
way one does not need to verify {\em a posteriori} that
$c_1,c_2,c_3$ are integrals of motion, because they are manifestly
even functions of $\epsilon$, while the symmetry of the linear
system with respect to $\epsilon$ has been broken.
\end{itemize}
To finish the proof along the lines of the first of the possible
arguments above, we show how to verify the statement that the
function $J$ in (\ref{eq: dC C1}) is an integral of motion, i.e.,
that
\[
\frac{p_1^2+p_2^2+p_3^2}
{1-\epsilon^2(\omega_1p_1^2+\omega_2p_2^2+\omega_3p_3^2)}=
\frac{\widetilde{p}_1^2+\widetilde{p}_2^2+\widetilde{p}_3^2}
{1-\epsilon^2(\omega_1\widetilde{p}_1^2+\omega_2\widetilde{p}_2^2+\omega_3
\widetilde{p}_3^2)}.
\]
This is equivalent to
\begin{eqnarray*}
\lefteqn{\widetilde{p}_1^{2}-p_1^2+\widetilde{p}_2^2-p_2^2+\widetilde{p}_3^2-p_3^2}\\
&=&\epsilon^2\left[(\omega_2-\omega_1)(\widetilde{p}_1^2p_2^2-\widetilde{p}_2^2p_1^2)
+(\omega_3-\omega_2)(\widetilde{p}_2^2p_3^2-\widetilde{p}_3^2p_2^2)
+(\omega_1-\omega_3)(\widetilde{p}_3^2p_1^2-\widetilde{p}_1^2p_3^2)\right].
\end{eqnarray*}
On the left-hand side of this equation we replace
$\widetilde{p}_i-p_i$ through the expressions from the last three
equations of motion (\ref{eq: dC}), on the right-hand side we
replace
$\epsilon(\omega_k-\omega_j)(\widetilde{p}_jp_k+p_j\widetilde{p}_k)$
by $\widetilde{m}_i-m_i$, according to the first three equations
of motion (\ref{eq: dC}). This brings the equation we want to
prove into the form
\begin{eqnarray*}
\lefteqn{(\widetilde{p}_1+p_1)(\widetilde{m}_3p_2+m_3\widetilde{p}_2-
\widetilde{m}_2p_3-m_2\widetilde{p}_3)\;+} & & \\
\lefteqn{(\widetilde{p}_2+p_2)(\widetilde{m}_1p_3+m_1\widetilde{p}_3-
\widetilde{m}_3p_1-m_3\widetilde{p}_1)\;+} & & \\
\lefteqn{(\widetilde{p}_3+p_3)(\widetilde{m}_2p_1+m_2\widetilde{p}_1-
\widetilde{m}_1p_2-m_1\widetilde{p}_2)\;=} & & \\
&&=(\widetilde{p}_1p_2-p_1\widetilde{p}_2)(\widetilde{m}_3-m_3)+
 (\widetilde{p}_2p_3-p_2\widetilde{p}_3)(\widetilde{m}_1-m_1)+
 (\widetilde{p}_3p_1-p_3\widetilde{p}_1)(\widetilde{m}_2-m_2).
\end{eqnarray*}

\noindent But the latter equation is an algebraic identity in
twelve variables $m_k,p_k,\widetilde{m}_k,\widetilde{p}_k$. This
finishes the proof. \hfill $\Box$

\paragraph{{\bf Remark}}
In the above proof and on many occasions below we make use of the
PSLQ algorithm in order to identify possible linear relations
among conserved quantities. Its applications are well documented
in the literature on Experimental Mathematics \cite{BB, BBG}, so
that we restrict ourselves here to a couple of minor remarks. We
apply the PSLQ algorithm to the numerical values of (the
candidates for) the conserved quantities obtained from the
algorithm (N). We note that it is crucial to apply the PSLQ
algorithm with many different initial data; from the large amount
of possible linear relations one should, of course, filter out
those relations which stay unaltered for different initial data.
It proved useful to perform these computations with rational data
(initial values of phase variables and parameters of the map) as
well as with high precision floating point numbers. In our
experiments we have been able to automate this task to a large
extent. All computations of this kind were performed on an Apple
MacBook with a 1.83 GHz Intel Core Duo processor and 2 GB of RAM.

\subsection{Remaining HK-bases}\label{Sect: dC Phi 123}
We now consider the remaining HK-bases $\Phi_1$,$\Phi_2$ and
$\Phi_3$. Here we are dealing with the three linear systems
\begin{eqnarray}
(\alpha_1p_1^2+\alpha_2p_2^2+\alpha_3p_3^2+\alpha_4 m_1^2
+\alpha_5 m_2^2+\alpha_6 m_3^2)\circ f^i(m,p) & = & m_1p_1\circ
f^i(m,p),
\label{eq: Psi1 iterates}\\
(\beta_1p_1^2+\beta_2p_2^2+\beta_3p_3^2+\beta_4m_1^2+\beta_5m_2^2
+\beta_6 m_3^2)\circ f^i(m,p) & = & m_2p_2\circ f^i(m,p),
\label{eq: Psi2 iterates}\\
(\gamma_1 p_1^2+\gamma_2p_2^2+\gamma_3p_3^2+\gamma_4m_1^2+\gamma_5
m_2^2+\gamma_6 m_3^2)\circ f^i(m,p) & = & m_3p_3\circ
f^i(m,p),\qquad \label{eq: Psi3 iterates}
\end{eqnarray}
already made non-homogeneous by normalizing the last coefficient
in each system, as in recipe (B), with $l=7$. The claim about each
of the systems is that it admits a unique solution for $i\in\bbZ$.
It is enough to solve each system for two different but
intersecting ranges of $l-1=6$ consecutive indices $i$, such as
$i\in[-2,3]$ and $i\in[-3,2]$, and to show that solutions coincide
for both ranges (recipe (C)). Actually, since the index range
$i\in[-2,3]$ is non-symmetric, it would be enough to consider the
system for this one range and to show that the solutions
$\alpha_j,\beta_j,\gamma_j$ are even functions with respect
to $\epsilon$ (recipe (D)). However, symbolic manipulations with
the iterates $f^i$ for $i=\pm 2,\pm 3$ are impossible. In what
follows, we will gradually extend the available information about
the coefficients $\alpha_j,\beta_j,\gamma_j$, which at the end
will allow us to get the analytic expressions for all of them and
to prove that they are integrals, indeed.

\subsection{First additional HK-basis}
\begin{figure}[hbt]
\centering \subfigure[] { \scalebox{0.5} {
\includegraphics{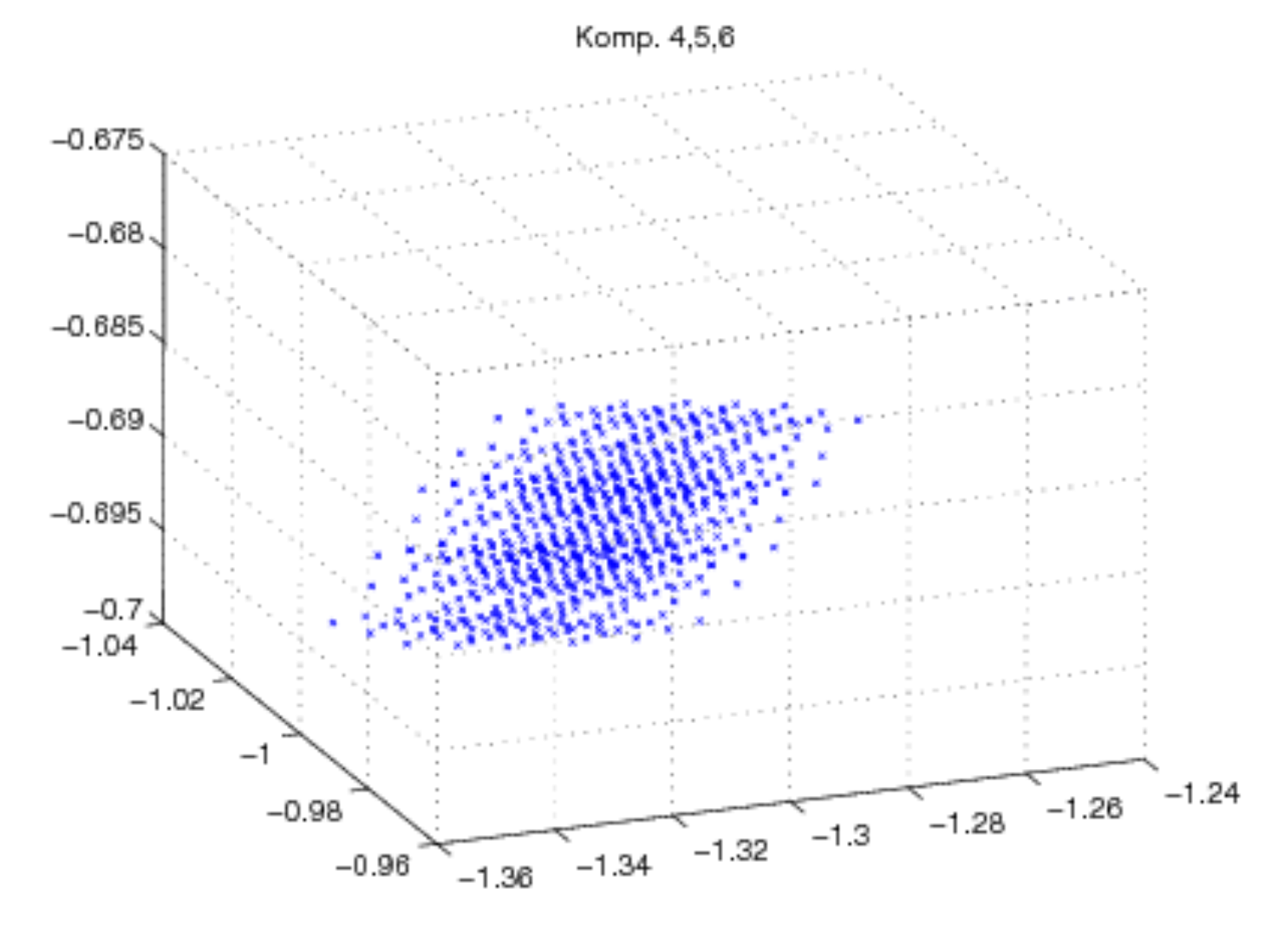} } } \subfigure[] {
\scalebox{0.5} { \includegraphics{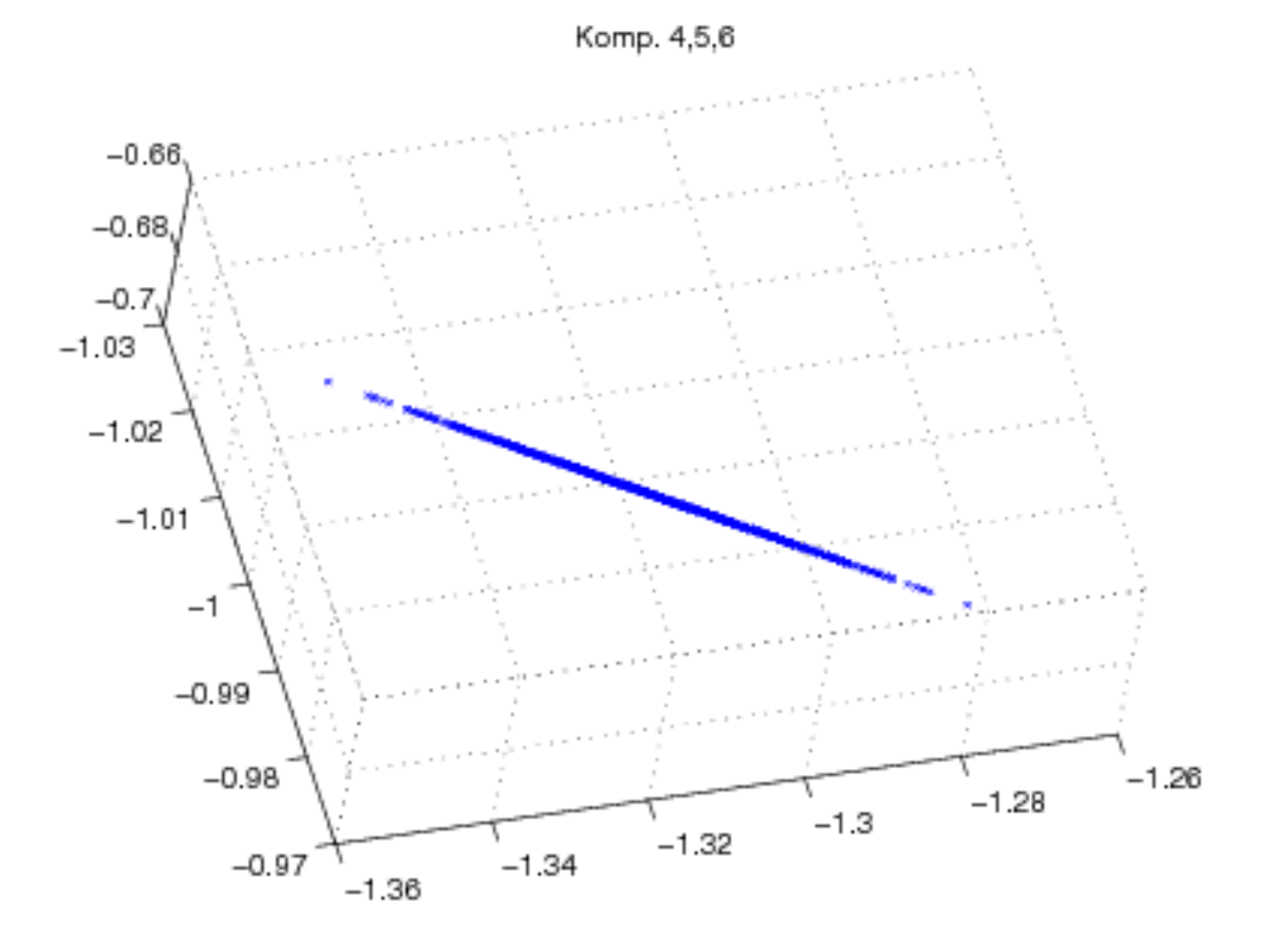} } }
\caption{Plot of the points $(d_7,d_8,d_9)$ for 729 values of
$(m,p)$ from a six-dimensional grid around the point
$(1,1,1,1,1,1)$ with a grid size of  0.01 and the parameters
$\epsilon=0.1$, $\omega_1=0.1$, $\omega_2=0.2$, $\omega_3=0.3$. }
\label{plot4}
\end{figure}

Theorem \ref{Th: dClebsch 1dim bases} shows that, after finding
the HK-basis $\Phi_0$ with $\dim K_{\Phi_0}(x)=1$ it is enough to
concentrate on (sub)-bases not containing the constant function
$\varphi_{10}(m,p)=1$. It turns out to be possible to find a
HK-basis without $\varphi_{10}$ and with a one-dimensional
null-space, which is more amenable to a symbolic treatment than
$\Phi_1,\Phi_2,\Phi_3$. Numerical algorithm (N) suggests that the
following set of functions is a HK-basis with $d=1$:
\begin{equation}\label{eq: dC basis 2}
\Psi=(p_1^2,p_2^2,p_3^2,m_1p_1,m_2p_2,m_3p_3).
\end{equation}
\begin{theorem}\label{Th: dC basis 2}
The set (\ref{eq: dC basis 2}) is a HK-basis for the map (\ref{eq:
dC map}) with $\dim K_{\Psi}(m,p)=1$. At every point
$(m,p)\in\bbR^6$ there holds:
\begin{equation}\nonumber
K_{\Psi}(m,p)=[-1:-1:-1:d_7:d_8:d_9],
\end{equation}
with
\begin{equation}
\label{eq: dC basis 2 ints}
d_k=\frac{(p_1^2+p_2^2+p_3^2)
(1+\epsilon^2d_k^{(2)}+\epsilon^4d_k^{(4)}+\epsilon^6d_k^{(6)})}
{\Delta}\,, \quad k=7,8,9,
\end{equation}
\begin{equation}
\label{eq: Delta}
\Delta=m_1p_1+m_2p_2+m_3p_3+\epsilon^2\Delta^{(4)}+\epsilon^4\Delta^{(6)}+
\epsilon^6\Delta^{(8)},
\end{equation}
where $d^{(2q)}_k$ and $\Delta^{(2q)}$ are homogeneous polynomials
of degree $2q$ in phase variables. In particular,
\begin{eqnarray}
d_7^{(2)} & = &
m_1^2+m_2^2+m_3^2+(\omega_2+\omega_3-2\omega_1)p_1^2+(\omega_3-\omega_2)p_2^2+
(\omega_2-\omega_3)p_3^2, \nonumber
\\
d_8^{(2)} & = &
m_1^2+m_2^2+m_3^2+(\omega_3-\omega_1)p_1^2+(\omega_3+\omega_1-2\omega_2)p_2^2+
(\omega_1-\omega_3)p_3^2, \nonumber
\\
d_9^{(2)} & = &
m_1^2+m_2^2+m_3^2+(\omega_2-\omega_1)p_1^2+(\omega_1-\omega_2)p_2^2+
(\omega_1+\omega_2-2\omega_3)p_3^2, \qquad \nonumber
\end{eqnarray}
and
\begin{equation}\nonumber
\Delta^{(4)}=m_1p_1d_7^{(2)}+m_2p_2d_8^{(2)}+m_3p_3d_9^{(2)}.
\end{equation}
(All other polynomials are too messy to be given here.) The
functions $d_7,d_8,d_9$ are integrals of the map (\ref{eq: dC
map}). They are dependent due to the linear relation
\begin{equation}
\label{eq: dC basis 2 lin rel}
(\omega_2-\omega_3)d_7+(\omega_3-\omega_1)d_8+
(\omega_1-\omega_2)d_9=0.
\end{equation}
Any two of them are functionally independent. Moreover, any two of
them together with $J$ are still functionally independent.
\end{theorem}
{\bf Proof.} As already mentioned, numerical experiments suggest
that for any $(m,p)\in\bbR^6$ there exists a one-dimensional space
of vectors $(d_1,d_2,d_3,d_7,d_8,d_9)$ satisfying
\begin{equation}\nonumber
(d_1p_1^2+d_2p_2^2+d_3p_3^2+d_7m_1p_1+d_8m_2p_2+d_9m_3p_3) \circ
f^i(m,p)=0
\end{equation}
for $i=0,1,\ldots,5$. According to recipe (A), one can equally
well consider this system for $i=-2,-1,\ldots,3$, which however
still contains the third iterate of $f$ and is therefore not
manageable. Therefore, we apply recipe (E) and look for linear
relations between the (numerical) solutions. Two such relations
can be observed immediately, namely
\begin{equation}
\label{eq: dC basis_2 rels 1}
d_1=d_2=d_3.
\end{equation}
Accepting these (still hypothetical) relations and applying recipe
(B), i.e., setting the common value of (\ref{eq: dC basis_2 rels
1}) equal to $-1$, we arrive at the non-homogeneous system of only
3 linear relations
\begin{equation}\label{eq: dC ansatz_2}
(d_7m_1p_1+d_8m_2p_2+d_9m_3p_3)\circ f^i(m,p)=
(p_1^2+p_2^2+p_3^2)\circ f^i(m,p)
\end{equation}
for $i=-1,0,1$. Fortunately, it is possible to find one more
linear relation between $d_7,d_8,d_9$. This was discovered
numerically: we produced a three-dimensional plot of the points
$(d_7(m,p),d_8(m,p),d_9(m,p))$ which can be seen in
Fig.~\ref{plot4} in two different projections.
This figure suggests that all these points lie on a plane in
$\bbR^3$, the second picture being a ``side view'' along a
direction parallel to this plane. Thus, it is plausible that one
more linear relation exists. With the help of the PSLQ algorithm
this hypothetic relation can be then identified as eq. (\ref{eq:
dC basis 2 lin rel}). Now the ansatz (\ref{eq: dC ansatz_2}) is
reduced to the following system of three equations for
$(d_7,d_8,d_9)$, which involves only one iterate of the map $f$:
\begin{equation}\label{eq: dC ansatz_2_final}
\renewcommand{\arraystretch}{1.3}
\left\{\begin{array}{l} (d_7m_1p_1+d_8m_2p_2+d_9m_3p_3)\circ
f^i(m,p) =(p_1^2+p_2^2+p_3^2)\circ f^i(m,p), \quad i=0,1, \\
(\omega_2-\omega_3)d_7+(\omega_3-\omega_1)d_8+(\omega_2-\omega_2)d_9
=0. \end{array}\right.
\end{equation}
This system can be solved by MAPLE, resulting in functions given
in eqs. (\ref{eq: dC basis 2 ints}), (\ref{eq: Delta}). These
(long) expressions can be found in \cite{WS}. They are manifestly
even functions of $\epsilon$, while the system has no symmetry
with respect to $\epsilon\mapsto-\epsilon$. This proves that they
are integrals of motion for the map $f$. This argument slightly
generalizes the recipes (D) and (E), and, since it is used not
only here but also on several further occasions in this paper, we
give here its formalization.
\begin{prop}\label{prop: even}
Consider a map $f:\bbR^6\to\bbR^6$ depending on a parameter
$\epsilon$, reversible in the sense of eq. (\ref{eq: dC reverse}).
Let $I(m,p,\epsilon)$ be an integral of $f$, even in $\epsilon$,
and let $A_1,A_2,A_3\in\bbR$. Suppose that the set of functions
$\Phi=(\varphi_1,\ldots,\varphi_4)$ is such that the system of
three linear equations for $(a_1,a_2,a_3)$,
\begin{equation}\label{eq: syst for even}
\renewcommand{\arraystretch}{1.3}
\left\{\begin{array}{l}
(a_1\varphi_1+a_2\varphi_2+a_3\varphi_3)\circ
f^i(m,p,\epsilon) =\varphi_4\circ f^i(m,p,\epsilon), \quad i=0,1, \\
A_1a_1+A_2a_2+A_3a_3 =I(m,p,\epsilon),
\end{array}\right.
\end{equation}
admits a unique solution which is even with respect to $\epsilon$.
Then this solution $(a_1,a_2,a_3)$ consists of integrals of the
map $f$, and $\Phi$ is a HK-basis with $\dim K_{\Phi}(m,p)=1$.
\end{prop}
{\bf Proof.} Since $(a_1,a_2,a_3)$ are even functions of
$\epsilon$, they satisfy also the system (\ref{eq: syst for even})
with $\epsilon\mapsto -\epsilon$, which, due to the reversibility
(\ref{eq: dC reverse}), can be represented as
\begin{equation}\label{eq: syst for even minus}
\left\{\begin{array}{l}
(a_1\varphi_1+a_2\varphi_2+a_3\varphi_3)\circ
f^i(m,p,\epsilon) =\varphi_4\circ f^i(m,p,\epsilon), \quad i=0,-1, \\
A_1a_1+A_2a_2+A_3a_3 =I(m,p,\epsilon).
\end{array}\right.
\end{equation}
Since the functions $(a_1,a_2,a_3)$ are {\em uniquely determined}
by any of the systems (\ref{eq: syst for even}) or (\ref{eq: syst
for even minus}), we conclude that they remain invariant under the
change $(m,p)\mapsto f(m,p,\epsilon)$, or, in other words, that
they are integrals of motion. Finally, we can conclude that these
functions satisfy equation
$(a_1\varphi_1+a_2\varphi_2+a_3\varphi_3)\circ f^i =\varphi_4\circ
f^i$  for all $i\in\bbZ$ (and can be uniquely determined by this
property), and that linear relation $A_1a_1+A_2a_2+A_3a_3=I$ is
satisfied, as well. \hfill $\Box$

Application of Proposition \ref{prop: even} to system (\ref{eq: dC
ansatz_2_final}) shows that $d_7,d_8,d_9$ are integrals of motion,
since they are even in $\epsilon$. Note that here, as always in
similar context, the evenness of solutions is due to ``miraculous
cancellation'' of the equal non-even polynomials which factor out
both in the numerators and denominators of the solutions. In the
present computation, these common non-even factors are of degree 2
in $\epsilon$.

It remains to prove that any two of the integrals $d_7,d_8,d_9$
together with the previously found integral $J$ are functionally
independent. For this aim, we show that from such a triple of
integrals one can construct another triple of integrals which
yields in the limit $\epsilon\to 0$ three independent conserved
quantities $H_3,H_4,H_1$ of the continuous Clebsch system. Indeed:
\begin{eqnarray*}
J & = & p_1^2+p_2^2+p_3^2+O(\epsilon^2)=H_3+O(\epsilon^2),\\
\frac{J}{d_{k+6}} & = &
m_1p_1+m_2p_2+m_3p_3+O(\epsilon^2)=H_4+O(\epsilon^2).
\end{eqnarray*}
On the other hand, it is easy to derive:
\[
\frac{d_7}{d_8}=1+\epsilon^2(d_7^{(2)}-d_8^{(2)})+O(\epsilon^4)=
1+\epsilon^2(\omega_2-\omega_1)(p_1^2+p_2^2+p_3^2)+O(\epsilon^4),
\]
and, taking this into account and computing the terms of order
$\epsilon^4$, one finds:
\[
\frac{d_7}{d_8}-1-\epsilon^2(\omega_2-\omega_1)J=
\epsilon^4(\omega_2-\omega_1)(2H_4^2+\omega_2H_3^2-2H_3H_1)+O(\epsilon^6),
\]
from which one easily extracts $H_1$. This proves our claim.
\hfill $\Box$

\paragraph{\bf Remark} With the basis $\Psi$, we encounter for the first
time the following interesting phenomenon: it can happen that a
HK-basis with a one-dimensional null-space provides several (in
this case two) functionally independent integrals. With Theorem
\ref{Th: dC basis 2}, we established existence of three
independent conserved quantities and two HK-bases with linearly
independent null-spaces. So, every orbit of the discrete Clebsch
system is shown to lie in a three-dimensional manifold which
belongs to an intersection of two quadrics in $\bbR^6$. The aim of
the following is to find one more independent integral and two
more HK-bases with one-dimensional null-spaces linearly
independent on $K_{\Phi_0}$, $K_\Psi$.

\subsection{Second additional HK-basis}
From the (still hypothetic) properties (\ref{eq: Psi1
iterates})--(\ref{eq: Psi3 iterates}) of the bases
$\Phi_1,\Phi_2,\Phi_3$ there follows that for any $(m,p)\in\bbR^6$
the system of linear equations
\begin{equation}\label{eq: Theta iterates}
(g_1p_1^2+g_2p_2^2+g_3p_3^2+g_4m_1^2+g_5m_2^2+g_6m_3^2)\circ
f^i(m,p)=(m_1p_1+m_2p_2+m_3p_3)\circ f^i(m,p)\;\;
\end{equation}
has a unique solution $(g_1,g_2,g_3,g_4,g_5,g_6)$. Indeed, the
solution should be given by
\begin{equation}
\label{eq: dC hs}
g_j=\alpha_j+\beta_j+\gamma_j,\quad j=1,\ldots,6.
\end{equation}
As for the bases $\Phi_1,\Phi_2,\Phi_3$, the solution of
(\ref{eq: Theta iterates}) can be determined by solving these
equations for two different but intersecting ranges of 6
consecutive values of $i$, say for $i\in[-3,2]$ and $i\in[-2,3]$.
However, it turns out that, due to the existence of several linear
relations between the solutions $g_j$, system (\ref{eq: Theta
iterates}) is much easier to deal with than systems (\ref{eq: Psi1
iterates})--(\ref{eq: Psi3 iterates}), so that the functions $g_j$
can be determined and studied independently of
$\alpha_j,\beta_j,\gamma_j$.
\begin{theorem}\label{Th: dC Theta}
The set of functions
\begin{equation}\nonumber
\Theta=(p_1^2,p_2^2,p_3^2,m_1^2,m_2^2,m_3^2,m_1p_1+m_2p_2+m_3p_3)
\end{equation}
is a HK-basis for the map (\ref{eq: dC map}) with $\dim
K_\Theta(m,p)=1$. At every point $(m,p)\in\bbR^6$ there holds:
\begin{equation}\nonumber
    K_\Theta(m,p)=[g_1:g_2:g_3:g_4:g_5:g_6:-1].
\end{equation}
Here $g_1,g_2,g_3$ are integrals of the map (\ref{eq: dC map})
given by
\begin{equation} \nonumber
g_k=\frac{g_k^{(4)}+\epsilon^2g_k^{(6)}+\epsilon^4g_k^{(8)}+
\epsilon^6g_k^{(10)}}{2(p_1^2+p_2^2+p_3^2)\Delta}\,,\qquad
k=1,2,3,
\end{equation}
where $g_k^{(2q)}$ are homogeneous polynomials of degree $2q$ in
phase variables, and $\Delta$ is given in eq. (\ref{eq: Delta}).
For instance,
\[
g_k^{(4)}=2H_4^2-H_3H_1+\omega_kH_3^2.
\]
Integrals $g_4,g_5,g_6$ are given by
\begin{equation}\nonumber
\label{eq: Theta rels}
g_4=\frac{g_2-g_3}{\omega_2-\omega_3}\,,\quad
g_5=\frac{g_3-g_1}{\omega_3-\omega_1}\,,\quad
g_6=\frac{g_1-g_2}{\omega_1-\omega_2}\,.
\end{equation}
\end{theorem}
{\bf Proof.} Since system (\ref{eq: Theta iterates}) involves too
many iterates of $f$ for a symbolical treatment, we look for
linear relations between the (numerical) solutions of this system.
Application of the PSLQ algorithm allows us to identify three such
relations, as given in eq. (\ref{eq: Theta rels}). This reduces
system (\ref{eq: Theta iterates}) to the following one:
\begin{eqnarray}\label{clebsch_ansatz_3a}
&&\hspace{-0.5cm}\bigg[
g_1\Big(p_1^2+\frac{m_2^2}{\omega_1-\omega_3}+\frac{m_3^2}{\omega_1-\omega_2}\Big)
+g_2\Big(p_2^2+\frac{m_1^2}{\omega_2-\omega_3}+\frac{m_3^2}{\omega_2-\omega_1}\Big)
\nonumber\\
&&+g_3\Big(p_3^2+\frac{m_1^2}{\omega_3-\omega_2}+\frac{m_2^2}{\omega_3-\omega_1}\Big)
\bigg]\circ f^i(m,p)  =  (m_1p_1+m_2p_2+m_3p_3)\circ
f^i(m,p).\qquad\quad
\end{eqnarray}
Thus, one can say that we are dealing with a reduced Hirota-Kimura
basis consisting of $l=4$ functions
\[
\widetilde{\Theta} = (I_1,I_2,I_3,H_4),
\]
see (\ref{eq: Clebsch I}). Interestingly, this is a basis of
integrals for the continuous-time Clebsch system, but we do not
know whether this is just a coincidence or has some deeper
meaning. System (\ref{clebsch_ansatz_3a}) has to be solved for two
different but intersecting ranges of $l-1=3$ consecutive indices
$i$. It would be enough to show that the solution for one
non-symmetric range, e.g., for $i\in[0,2]$, consists of even
functions of $\epsilon$. However, this non-symmetric system
involves with necessity the second iterate $f^2$. To avoid dealing
with $f^2$, one more linear relation for $g_1,g_2,g_3$ would be
needed. Such a relation has been found with the help of PSLQ
algorithm, it does not have constant coefficients anymore but
involves the previously found integrals $d_7,d_8,d_9$:
\begin{equation}\label{eq: dC Theta additional}
(\omega_2-\omega_3)g_1+(\omega_3-\omega_1)g_2+(\omega_1-\omega_2)g_3=
\frac{1}{2}(\omega_2-\omega_3)(\omega_3-\omega_1)(d_8-d_7).
\end{equation}
Of course, due to eq. (\ref{eq: dC basis 2 lin rel}), the
right-hand side of eq. (\ref{eq: dC Theta additional}) can be
equivalently put as
\[
\frac{1}{2}(\omega_3-\omega_1)(\omega_1-\omega_2)(d_9-d_8)=
\frac{1}{2}(\omega_1-\omega_2)(\omega_2-\omega_3)(d_7-d_9).
\]
The linear system consisting of eq. (\ref{clebsch_ansatz_3a}) for
$i=0,1$ and eq. (\ref{eq: dC Theta additional}) can be solved by
MAPLE with the result given in theorem. Since $(d_7,d_8,d_9)$ are
already proven to be integrals of motion, and since the solutions
$(g_1,g_2,g_3)$ are manifestly even in $\epsilon$, Proposition
\ref{prop: even} yields that $(g_1,g_2,g_3)$ are integrals of the
map $f$. \hfill $\Box$

Theorem \ref{Th: dC Theta} gives us the third HK-basis with a
one-dimensional null-space for the discrete Clebsch system. Thus,
it shows that every orbit lies in the intersection of three
quadrics in $\bbR^6$. What concerns the integrals of motion, it
turns out that the basis $\Theta$ does not provide us with
additional ones: a numerical check with gradients shows that
integrals $g_1,g_2,g_3$ are functionally dependent from the
previously found ones. At this point we are lacking one more
HK-basis with a one-dimensional null-space, linearly independent
from $K_{\Phi_0}$, $K_\Psi$, $K_\Theta$, and one more integral of
motion, functionally independent from $J$ and $d_7,d_8$.

\subsection{Proof for the bases $\Phi_1,\Phi_2,\Phi_3$}
Now we return to the bases $\Phi_1,\Phi_2,\Phi_3$ discussed in
Sect. \ref{Sect: dC Phi 123}. In order to be able to solve systems
(\ref{eq: Psi1 iterates})--(\ref{eq: Psi3 iterates}) symbolically
and to prove that the solutions $\alpha_j,\beta_j,\gamma_j$ are
indeed integrals, we have to find additional linear relations for
these quantities (recipe (E)). Within each set of coefficients we
were able to identify just one relation:
\begin{eqnarray}
(\omega_1-\omega_3)\alpha_5 & = & (\omega_1-\omega_2)\alpha_6,
\label{eq: rel18_1}\\
(\omega_2-\omega_3)\beta_4 & = & (\omega_2-\omega_1)\beta_6,
\label{eq: rel18_2} \\
(\omega_3-\omega_2)\gamma_4 & = & (\omega_3-\omega_1)\gamma_5.
\label{eq: rel18_3}
\end{eqnarray}
This reduces the number of equations in each system by one, which
however does not resolve our problems. A way out consists in
looking for linear relations among all the coefficients
$\alpha_j, \beta_j, \gamma_j$. Remarkably, six more independent
linear relations of this kind can be identified:
\begin{equation}\label{eq: rel18_4}
\alpha_4=\beta_5=\gamma_6,
\end{equation}
\begin{eqnarray}
&& \frac{\alpha_2-\alpha_3-(\omega_2-\omega_3)\alpha_4}{\omega_2-\omega_3}=
\frac{\beta_2-\beta_3-(\omega_2-\omega_3)\beta_4}{\omega_3-\omega_1}
=
\frac{\gamma_2-\gamma_3-(\omega_2-\omega_3)\gamma_4}{\omega_1-\omega_2},
\qquad\label{eq: rel18_6}\\
&&\frac{\alpha_3-\alpha_1-(\omega_3-\omega_1)\alpha_5}{\omega_2-\omega_3}=
\frac{\beta_3-\beta_1-(\omega_3-\omega_1)\beta_5}{\omega_3-\omega_1}
=
\frac{\gamma_3-\gamma_1-(\omega_3-\omega_1)\gamma_5}{\omega_1-\omega_2}.
\label{eq: rel18_8}
\end{eqnarray}
There are two more similar relations:
\beq \nonumber
\frac{\alpha_1-\alpha_2-(\omega_1-\omega_2)\alpha_6}{\omega_2-\omega_3}=
\frac{\beta_1-\beta_2-(\omega_1-\omega_2)\beta_6}{\omega_3-\omega_1}=
\frac{\gamma_1-\gamma_2-(\omega_1-\omega_2)\gamma_6}{\omega_1-\omega_2},
\eeq
but they follow from the already listed ones (\ref{eq:
rel18_1})--(\ref{eq: rel18_8}). We stress that all these linear
relations were identified numerically, with the help of the PSLQ
algorithm, and remain at this stage hypothetic.

With nine linear relations (\ref{eq: rel18_1})--(\ref{eq:
rel18_8}), we have to solve systems (\ref{eq: Psi1
iterates})--(\ref{eq: Psi3 iterates}) {\em simultaneously} for a
range of 3 consecutive indices $i$. Taking this range as
$i=-1,0,1$ we can avoid dealing with $f^2$, which however would
leave us with the problem of a proof that the solutions are
integrals. Alternatively, we can choose the range $i=0,1,2$, and
then the solutions are automatically integrals, as soon as it is
established that they are even functions of $\epsilon$.

A symbolic solution of the system consisting of 18 linear
equations, namely eqs. (\ref{eq: Psi1 iterates})--(\ref{eq: Psi3
iterates}) with $i=0,1,2$ along with nine simple equations
(\ref{eq: rel18_1})--(\ref{eq: rel18_8}), would require
astronomical amounts of memory, because of the complexity of
$f^2$. However, this task becomes manageable and even simple for
fixed (numerical) values of the phase variables $(m,p)$ and of the
parameters $\omega_i$, while leaving $\epsilon$ a symbolic
variable. For rational values of $m_k,p_k,\omega_k$ all
computations can be done precisely (in rational arithmetic). This
means that $\alpha_j$, $\beta_j$, and $\gamma_j$ can be evaluated,
as functions of $\epsilon$, at arbitrary points in
$\bbQ^9(m,p,\omega_1,\omega_2,\omega_3)$. A big number of such
evaluations provides us with a convincing evidence in favor of the
claim that these functions are even in $\epsilon$.

In order to obtain a rigorous proof without dealing with $f^2$,
further linear relations would be necessary. Before introducing
these, we present some preliminary considerations. Assuming that
$\Phi_1,\Phi_2,\Phi_3$ are HK-bases with one-dimensional
null-spaces, results of Theorem \ref{Th: dC basis 2} on the
HK-basis $\Psi$ tell us that the row vector $(d_7,d_8,d_9)$ is the
unique left null-vector for the matrix
\[
M_2=\begin{pmatrix} \alpha_4 & \alpha_5 & \alpha_6 \\
                \beta_4  & \beta_5  & \beta_6  \\
                \gamma_4 & \gamma_5 & \gamma_6
\end{pmatrix},
\]
normalized so that
\[
(d_7,d_8,d_9)M_1=(1,1,1),\quad {\rm where}\quad
M_1=\begin{pmatrix} \alpha_1 & \alpha_2 & \alpha_3 \\
                    \beta_1  & \beta_2  & \beta_3  \\
                    \gamma_1 & \gamma_2 & \gamma_3
\end{pmatrix}.
\]
Note that due to eqs. (\ref{eq: rel18_1})--(\ref{eq: rel18_4}) the
matrix $M_2$ has at most four (linearly) independent entries.
Denoting the common values in these equations by $A,B,C,D$,
respectively, we find:
\begin{equation}
\label{eq: M2}
M_2=\begin{pmatrix} \alpha_4 & \alpha_5 & \alpha_6 \\
                \beta_4  & \beta_5  & \beta_6  \\
                \gamma_4 & \gamma_5 & \gamma_6
\end{pmatrix}
=\begin{pmatrix} D & A/(\omega_1-\omega_3) & A/(\omega_1-\omega_2) \\
                B/(\omega_2-\omega_3)  & D  & B/(\omega_2-\omega_1)  \\
                C/(\omega_3-\omega_2) & C/(\omega_3-\omega_1) & D
\end{pmatrix}.
\end{equation}
The existence of the left null-vector $(d_7,d_8,d_9)$ shows that
$\det(M_2)=0$, or, equivalently,
\begin{equation}\label{eq: ABCD dependence}
D^2-\frac{AB}{(\omega_1-\omega_3)(\omega_2-\omega_3)}
-\frac{BC}{(\omega_2-\omega_1)(\omega_3-\omega_1)}
-\frac{CA}{(\omega_3-\omega_2)(\omega_1-\omega_2)}=0.
\end{equation}
From eqs. (\ref{eq: M2}) and (\ref{eq: ABCD dependence}) one easily
derives that the row
\begin{eqnarray*}
&&\left(D-\frac{B}{\omega_2-\omega_3}-\frac{C}{\omega_3-\omega_2}\,,\,
D-\frac{A}{\omega_1-\omega_3}-\frac{C}{\omega_3-\omega_1}\,,\,
D-\frac{A}{\omega_1-\omega_2}-\frac{B}{\omega_2-\omega_1}\right)\\\\
&&
\qquad=(\alpha_4-\beta_4-\gamma_4,\,-\alpha_5+\beta_5-\gamma_5,\,
-\alpha_6-\beta_6+\gamma_6)
\end{eqnarray*}
is a left null-vector of the matrix $M_2$, and therefore
$(d_7,d_8,d_9)$ is proportional to this vector. The
proportionality coefficient can be now determined with the help of
the PSLQ algorithm and turns out to be extremely simple. Namely,
the following relations hold:
\begin{eqnarray}
\alpha_4-\beta_4-\gamma_4 & = &
D-\frac{B-C}{\omega_2-\omega_3}\;=\;\frac{1}{2}\,d_7,
 \label{eq: dC Psi add lin rel 1}\\
-\alpha_5+\beta_5-\gamma_5 & = &
D-\frac{C-A}{\omega_3-\omega_1}\;=\;\frac{1}{2}\,d_8,
 \label{eq: dC Psi add lin rel 2}\\
-\alpha_6-\beta_6+\gamma_6 & = &
D-\frac{A-B}{\omega_1-\omega_2}\;=\;\frac{1}{2}\,d_9.
 \label{eq: dC Psi add lin rel 3}
\end{eqnarray}
Only two of them are independent, because of eq. (\ref{eq: dC
basis 2 lin rel}). We note also that, according to eq. (\ref{eq:
dC hs}), one has
\begin{eqnarray}
\alpha_4+\beta_4+\gamma_4 & = &
D+\frac{B-C}{\omega_2-\omega_3}\;=\;g_4,
 \label{eq: dC Psi add lin rel 4}\\
\alpha_5+\beta_5+\gamma_5 & = &
D+\frac{C-A}{\omega_3-\omega_1}\;=\;g_5,
 \label{eq: dC Psi add lin rel 5}\\
\alpha_6+\beta_6+\gamma_6 & = &
D+\frac{A-B}{\omega_1-\omega_2}\;=\;g_6.
 \label{eq: dC Psi add lin rel 6}
\end{eqnarray}
Equations (\ref{eq: dC Psi add lin rel 1})--(\ref{eq: dC Psi add
lin rel 6}) and (\ref{eq: ABCD dependence}) are already enough to
determine all four integrals $A,B,C,D$, that is, all
$\alpha_j,\beta_j,\gamma_j$ with $j=4,5,6$, {\em provided} it is
proven that they are indeed integrals. These (conditional) results
read:
\begin{eqnarray}
A & = &
\frac{1+\epsilon^2A^{(2)}+\epsilon^4A^{(4)}+\epsilon^6A^{(6)}+\epsilon^8A^{(8)}}
{2\epsilon^2\Delta}\,,  \label{eq: dC A}\\
B & = &
\frac{1+\epsilon^2B^{(2)}+\epsilon^4B^{(4)}+\epsilon^6B^{(6)}+\epsilon^8B^{(8)}}
{2\epsilon^2\Delta}\,,  \label{eq: dC B}\\
C & = &
\frac{1+\epsilon^2C^{(2)}+\epsilon^4C^{(4)}+\epsilon^6C^{(6)}+\epsilon^8C^{(8)}}
{2\epsilon^2\Delta}\,,  \label{eq: dC C}\\
D & = &
\frac{p_1^2+p_2^2+p_3^2+\epsilon^2D^{(4)}+\epsilon^4D^{(6)}+\epsilon^6D^{(8)}}
{2\Delta}\,,  \label{eq: dC D}
\end{eqnarray}
where $A^{(2q)}$, $B^{(2q)}$, $C^{(2q)}$, $D^{(2q)}$ are
homogeneous polynomials of degree $2q$ in phase variables, for
instance,
\begin{eqnarray*}
A^{(2)} & = &
B^{(2)}\;=\;C^{(2)}\\
& = & m_1^2+m_2^2+m_3^2+(\omega_2+\omega_3-2\omega_1)p_1^2+
(\omega_3+\omega_1-2\omega_2)p_2^2+(\omega_1+\omega_2-2\omega_3)p_3^2,\\
D^{(4)} & = &
(m_1p_1+m_2p_2+m_3p_3)^2\\
&& +(p_1^2+p_2^2+p_3^2)\Big((\omega_2+\omega_3-2\omega_1)p_1^2+
(\omega_3+\omega_1-2\omega_2)p_2^2+(\omega_1+\omega_2-2\omega_3)p_3^2\Big).
\end{eqnarray*}
We remark that eq. (\ref{eq: ABCD dependence}) tells us that no
more than three of the functions $A,B,C,D$ are actually
functionally independent. Computation with gradients shows that
$A,B,C$ are functionally independent, indeed. Moreover, all other
previously found integrals $J$, $d_7,d_8,d_9$, and $g_1,g_2,g_3$
are functionally dependent on these ones.

\begin{theorem}\label{Th: dC bases 345}
The sets (\ref{eq: dC basis Psi1})--(\ref{eq: dC basis Psi3}) are
HK-bases for the map (\ref{eq: dC map}) with $\dim
K_{\Phi_1}(m,p)=\dim K_{\Phi_2}(m,p)=\dim K_{\Phi_3}(m,p)=1$. At
each point $(m,p)\in\bbR^6$ there holds:
\begin{eqnarray}
K_{\Phi_1}(m,p) & = &
[\alpha_1:\alpha_2:\alpha_3:\alpha_4:\alpha_5:\alpha_6:-1], \nonumber \\
K_{\Phi_2}(m,p) & = &
[\beta_1:\beta_2:\beta_3:\beta_4:\beta_5:\beta_6:-1],\nonumber\\
K_{\Phi_3}(m,p) & = &
[\gamma_1:\gamma_2:\gamma_3:\gamma_4:\gamma_5:\gamma_6:-1],\nonumber
\end{eqnarray}
where $\alpha_j$,$\beta_j$, and $\gamma_j$ are rational functions
of $(m,p)$, even with respect to $\epsilon$. They are integrals of
motion for the map (\ref{eq: dC map}) and satisfy linear relations
(\ref{eq: rel18_1})--(\ref{eq: rel18_8}). For $j=4,5,6$ they are
given by eqs. (\ref{eq: M2}), (\ref{eq: dC C}), (\ref{eq: dC D}).
For $j=1,2,3$ they are of the form
\begin{equation} \label{eq: bigint_form}
h=\frac{h^{(2)}+\epsilon^2h^{(4)}+\epsilon^4h^{(6)}+
\epsilon^6h^{(8)}+ \epsilon^8h^{(10)}+\epsilon^{10}h^{(12)}}
{2\epsilon^2(p_1^2+p_2^2+p_3^2)\Delta}\,,
\end{equation}
where $h$ stands for any of the functions
$\alpha_j,\beta_j,\gamma_j$, $j=1,2,3$, and the corresponding
$h^{(2q)}$ are homogeneous polynomials in phase variables of
degree $2q$. For instance,
\begin{equation}\label{alpha123}
\renewcommand{\arraystretch}{1.2}
\begin{array}{lll}
\alpha_1^{(2)}=H_3-I_1, \quad & \alpha_2^{(2)}=-I_1, &
\alpha_3^{(2)}=-I_1,\\
\beta_1^{(2)}=-I_2, & \beta_2^{(2)}=H_3-I_2,  \quad  &
\beta_3^{(2)}=-I_2,\\
\gamma_1^{(2)}=-I_3, & \gamma_2^{(2)}=-I_3, &
\gamma_3^{(2)}=H_3-I_3.
\end{array}
\end{equation}
The four functions $J$, $\alpha_1$, $\beta_1$ and $\gamma_1$ are
functionally independent.
\end{theorem}
{\bf Proof.} The proof consists of several steps.

{\em Step 1.} Consider the system for 18 unknowns
$\alpha_j,\beta_j,\gamma_j$, $j=1,\ldots,6$, consisting of 17
linear equations: eqs. (\ref{eq: Psi1 iterates})--(\ref{eq: Psi3
iterates}) with $i=0,1$, eqs. (\ref{eq: rel18_1})--(\ref{eq:
rel18_8}), and eqs. (\ref{eq: dC Psi add lin rel 1}), (\ref{eq: dC
Psi add lin rel 2}). This system is underdetermined, so that in
principle it admits a one-parameter family of solutions.
Remarkably, the symbolic MAPLE solution shows that all variables
$\alpha_j,\beta_j,\gamma_j$ with $j=4,5,6$ are determined by this
system uniquely, the results coinciding with eqs. (\ref{eq: M2}),
(\ref{eq: dC C})--(\ref{eq: dC D}). (Actually, the MAPLE answers
are much more complicated, and their simplification has been
performed with SINGULAR, which was used to cancel out common
factors from the huge expressions in numerators and denominators
of these rational functions.) Since these uniquely determined
$\alpha_j,\beta_j,\gamma_j$ with $j=4,5,6$ are even functions of
$\epsilon$, this proves that they (i.e., $A,B,C,D$) are integrals
of motion.

{\em Step 2.} Having determined $\alpha_j,\beta_j,\gamma_j$ with
$j=4,5,6$, we are in a position to {\em compute}
$\alpha_j,\beta_j,\gamma_j$ with $j=1,2,3$. For instance, to
obtain the values of $\alpha_j$ with $j=1,2,3$, we consider the
symmetric linear system (\ref{eq: Psi1 iterates}) with $i=-1,0,1$
(and with already found $\alpha_4,\alpha_5,\alpha_6$). This system
has been solved by MAPLE. The solutions are huge rational
functions which however turn out to admit massive cancellations.
These cancellations have been performed with the help of SINGULAR.
The resulting expressions for $\alpha_1,\alpha_2,\alpha_3$ turn
out to satisfy the ansatz (\ref{eq: bigint_form}) with the leading
terms given in the first line of eq. (\ref{alpha123}). (All
further terms can be found in \cite{WS}.) However, this
computation does not prove that the functions so obtained are
indeed integrals of motion. To prove this, one could, in
principle, either check directly the identities $\alpha_j\circ
f=\alpha_j$, $j=1,2,3$, or verify equation (\ref{eq: Psi1
iterates}) with $i=2$. Both ways are prohibitively expensive, so
that we have to look for an alternative one.

{\em Step 3.} The results of Step 2 yield an explicit expression
for the function
\begin{equation}\label{eq:F}
F=(\omega_2-\omega_3)\alpha_1+(\omega_3-\omega_1)\alpha_2+(\omega_1-\omega_2)\alpha_3,
\end{equation}
which is of the form
\[
F=\frac{(\omega_2-\omega_3)(1+\epsilon^2F^{(2)}+\epsilon^4F^{(4)}+\epsilon^6F^{(6)}
+\epsilon^8F^{(8)})}{2\epsilon^2\Delta}.
\]
It is of a crucial importance for our purposes that it can be
proven directly that $F$ is an integral of motion. We have proved
this with the method (G) based on the Gr\"obner basis for the
ideal generated by discrete equations of motion. The application
of this method to $F$ is more feasible that to any single of
$\alpha_j$, $j=1,2,3$, because of the cancellation of the huge
polynomial coefficient of $\epsilon^{10}$ in the numerator of $F$.
Actually, more is true: $F$ is not only an integral, but is
functionally dependent on the previously found ones, say on
$J,d_7,d_8$. For a proof of this claim, it would be most favorable
to find the explicit dependence $F=F(J,d_7,d_8)$, but it remains
unknown to us. Instead, we have chosen the way of verification
that
\[
\nabla F\in{\rm span}(\nabla J,\nabla d_7,\nabla d_8).
\]
This is easily checked numerically for arbitrarily many (rational)
values of the data involved. For a symbolic check, one has to
prove the existence of three scalar functions
$\lambda_1,\lambda_2,\lambda_3$ such that
\[
\nabla F=\lambda_1\nabla J+\lambda_2\nabla d_7+\lambda_3\nabla
d_8.
\]
This is the system of six equations for three unknowns. Since $J$
does not depend on $m_k$, one can determine $\lambda_2$,
$\lambda_3$ from a system of only three equations:
\[
\nabla_m F=\lambda_2\nabla_m d_7+\lambda_3\nabla_m d_8.
\]
After that, it remains to check that $\nabla_p F-\lambda_2\nabla_p
d_7-\lambda_3\nabla_p d_8$ is proportional to $\nabla_pJ$.
Clearly, these computations can be arranged so as to verify
vanishing of certain (very big) polynomials. We have been able to
perform these computations with the help of SINGULAR for symbolic
$m_k,p_k$ but with (several sets of) numeric values of
coefficients $\omega_k$ only.

{\em Step 4.} The result of Step 3 allows us to proceed as
follows. Consider the system of three linear equations for
$\alpha_1,\alpha_2,\alpha_3$, consisting of (\ref{eq: Psi1
iterates}) with $i=0,1$, and of
\[
(\omega_2-\omega_3)\alpha_1+(\omega_3-\omega_1)\alpha_2+(\omega_1-\omega_2)\alpha_3=F,
\]
where $F$ is the explicit expression obtained and proven to be an
integral on Step 3. This system can now  be solved by MAPLE; the
results, again simplified with SINGULAR, are even functions of
$\epsilon$ (actually, the same ones obtained on Step 1 from the
symmetric system). Non-even polynomials in $\epsilon$ of degree 7
cancel in a miraculous way from the numerators and the
denominator. Now Proposition \ref{prop: even} assures that these
solutions are integrals of motion.

{\em Step 5.} Finally, in order to find $\beta_1,\beta_2,\beta_3$
and $\gamma_1,\gamma_2,\gamma_3$, we solve the two systems
consisting of (\ref{eq: Psi2 iterates}), resp. (\ref{eq: Psi3
iterates}) with $i=0,1$, and the first, resp. the second linear
relation in eq. (\ref{eq: rel18_6}). The results are even
functions of $\epsilon$, satisfying the ansatz (\ref{eq:
bigint_form}) with the leading terms given in eq.
(\ref{alpha123}). Proposition \ref{prop: even} yields that also
these functions are integrals of motion. \hfill $\Box$

MAPLE worksheets for all computations used in this section can be
found in \cite{WS}.

\subsection{Preliminary results on the Hirota-Kimura-type discretization
of the general flow of the Clebsch system}

The general flow of the Clebsch system, depending on three real
parameters $b_1,b_2,b_3$ (or, rather, on their differences
$b_i-b_j$, which gives two independent real parameters), reads as
follows:
\begin{equation}\label{gcl}
\left\{ \begin{array}{l}
\dot{m} =  m\times Cm+p\times Bp\,, \vspace{.2truecm}\\
\dot{p} =  p\times Cm,
\end{array} \right.
\end{equation}
where $B={\rm{diag}}(b_1,b_2,b_3)$ and
$C={\rm{diag}}(c_1,c_2,c_3)$ with
\begin{equation}
\label{eq: c thru b}
c_1=\frac{b_2-b_3}{\omega_2-\omega_3}\,,\quad
c_2=\frac{b_3-b_1}{\omega_3-\omega_1}\,,\quad
c_3=\frac{b_1-b_2}{\omega_1-\omega_2}\,.
\end{equation}
This flow is Hamiltonian with the quadratic Hamilton function
\[
H = \frac{1}{2}\,\langle m, Cm\rangle + \frac{1}{2}\,\langle p,
Bp\rangle=\frac{1}{2}\sum_{k=1}^3(c_km_k^2+b_kp_k^2)=
\frac{1}{2}(b_1I_1+b_2I_2+b_3I_3).
\]
In components eqs. (\ref{gcl}) read:
\begin{eqnarray}
\dot{m}_1 & = & (c_3-c_2)m_2m_3+(b_3-b_2)p_2p_3,  \nonumber\\
\dot{m}_2 & = & (c_1-c_3)m_3m_1+(b_1-b_3)p_3p_1,  \nonumber \\
\dot{m}_3 & = & (c_2-c_1)m_1m_2+(b_2-b_1)p_1p_2.  \nonumber \\
\dot{p}_1 & = & c_3m_3p_2-c_2m_2p_3, \nonumber  \\
\dot{p}_2 & = & c_1m_1p_3-c_3m_3p_1, \nonumber \\
\dot{p}_3 & = & c_2m_2p_1-c_1m_1p_2.\nonumber
\end{eqnarray}
The KH discretization of the flow (\ref{gcl}) reads
\begin{equation}\nonumber
 \left\{
\begin{array}{rcl}
\widetilde{m}-m & = & \ep(\widetilde{m} \times Cm + m \times
C\widetilde{m} + \widetilde{p} \times Bp+p\times
B\widetilde{p}\,),
\vspace{.2truecm}\\
\widetilde{p}-p & = & \ep\left(  \widetilde{p}\times Cm+p\times C
\widetilde{m} \right).
\end{array}
\right.
\end{equation}
In components:
\begin{eqnarray} \label{eq:dC2}
\widetilde{m}_1-m_1 &=&
\epsilon(c_3-c_2)(\widetilde{m}_2m_3+m_2\widetilde{m}_3)+
\epsilon(b_3-b_2)(\widetilde{p}_2p_3+p_2\widetilde{p}_3),
\nonumber \\
\widetilde{m}_2-m_2 &=&
\epsilon(c_1-c_3)(\widetilde{m}_3m_1+m_3\widetilde{m}_1)+
\epsilon(b_1-b_3)(\widetilde{p}_3p_1+p_3\widetilde{p}_1),
\nonumber \\
\widetilde{m}_3-m_3 &=&
\epsilon(c_2-c_1)(\widetilde{m}_1m_2+m_1\widetilde{m}_2)+
\epsilon(b_2-b_1)(\widetilde{p}_1p_2+p_1\widetilde{p}_2),
\nonumber \\
\widetilde{p}_1-p_1 &=& \epsilon
c_3(\widetilde{m}_3p_2+m_3\widetilde{p}_2)-\epsilon
c_2(\widetilde{m}_2 p_3+m_2\widetilde{p}_3),
\nonumber \\
\widetilde{p}_2-p_2 &=& \epsilon
c_1(\widetilde{m}_1p_3+m_1\widetilde{p}_3)-\epsilon
c_3(\widetilde{m}_3p_1+m_3\widetilde{p}_1),
\nonumber \\
\widetilde{p}_3-p_3 &=& \epsilon
c_2(\widetilde{m}_2p_1+m_2\widetilde{p}_1)-\epsilon
c_1(\widetilde{m}_1 p_2+m_1\widetilde{p}_2).
\end{eqnarray}
In what follows, we will use the abbreviations $b_{ij}=b_i-b_j$
and $c_{ij}=c_i-c_j$. The linear system (\ref{eq:dC2}) defines an
explicit, birational map $f:\bbR^6 \rightarrow \bbR^6$,
\begin{equation}\label{eq:dC2 map}
\begin{pmatrix}  \widetilde{m} \\ \widetilde{p} \end{pmatrix} = f(m,p,\epsilon) =
M^{-1}(m,p,\epsilon)\begin{pmatrix} m \\ p \end{pmatrix},
\end{equation}
where
\begin{equation}\nonumber
M(m,p,\epsilon) = \begin{pmatrix}
1 & \epsilon c_{23}m_3 & \epsilon c_{23}m_2 & 0 & \epsilon b_{23}p_3 & \epsilon b_{23}p_2 \\
\epsilon c_{31}m_3 & 1 & \epsilon c_{31}m_1 & \epsilon b_{31}p_3 & 0 & \epsilon b_{31}p_1 \\
\epsilon c_{12}m_2 & \epsilon c_{12}m_1 & 1 & \epsilon b_{12}p_2 & \epsilon b_{12}p_1 & 0 \\
0 & \epsilon c_2p_3 & -\epsilon c_3p_2 & 1 & -\epsilon c_3m_3 & \epsilon c_2m_2 \\
-\epsilon c_1p_3 & 0 & \epsilon c_3p_1 & \epsilon c_3m_3 & 1 & -\epsilon c_1m_1 \\
\epsilon c_1p_2 & -\epsilon c_2p_1 & 0 & -\epsilon c_2m_2 & \epsilon c_1m_1 & 1 \\
\end{pmatrix}.
\end{equation}
As usual, map (\ref{eq:dC2 map}) possesses the reversibility
property
\begin{equation}\nonumber
f^{-1}(m,p,\epsilon) = f(m,p,-\epsilon).
\end{equation}

\begin{conjecture} All claims of Theorems \ref{Th: dClebsch max
basis}, \ref{Th: dClebsch 1dim bases} hold also for the
discretization (\ref{eq:dC2 map}) of the general flow of the
Clebsch system, with the HK-basis $\Phi_0$ being replaced by
\begin{equation}\label{eq: Clebsch2 first basis}
\Phi_0=(p_1^2,p_2^2,p_3^2,m_1^2,m_2^2,m_3^2,1).
\end{equation}
\end{conjecture}
This conjecture is supported by numerical results based on the
algorithm (N). The claim concerning the set $\Phi_0$ given in eq.
(\ref{eq: Clebsch2 first basis}) is proven symbolically. In order
to keep the notations compact, we give here this proof for the
{\em second flow} of the Clebsch system only.

Recall that the first flow of the Clebsch system, considered in
Sect. \ref{Sect: Clebsch1}, corresponds to $b_i=\omega_i$ and
$c_i=1$. The second flow is characterized by the Hamilton function
\begin{equation}\nonumber
H=\frac{1}{2}H_2=\frac{1}{2}(\omega_1m_1^2+\omega_2m_2^2+\omega_3m_3^2
-\omega_2\omega_3p_1^2-\omega_1\omega_3p_2^2-\omega_1\omega_2p_3^2).
\end{equation}
In other words, the choice of parameters $b_k$ characterizing the
second flow is
\begin{equation}\label{eq: Clebsch 2 b}
b_1=-\omega_2\omega_3,\quad b_2=-\omega_3\omega_1,\quad
b_3=-\omega_1\omega_2,
\end{equation}
so that
\begin{equation}\label{eq: Clebsch 2 c}
c_1=\omega_1,\quad c_2=\omega_2,\quad c_3=\omega_3.
\end{equation}
For the HK discretization of the second Clebsch flow, we give a
more concrete formulation of our findings concerning the HK-basis
$\Phi_0$, including a ``nice'' integral.

\begin{theorem}
For the map (\ref{eq:dC2 map}) the set of functions (\ref{eq:
Clebsch2 first basis}) is a HK-basis with $\dim
K_{\Phi_0}(m,p)=1$. At each point $(m,p)\in\bbR^6$ there holds:
\[
K_{\Phi_0}(m,p)=[e_1:e_2:e_3:e_4:e_5:e_6:-1],
\]
where all $e_i$ are fractional-linear functions of a single
integral $L=L(m,p,\epsilon)$ of the map (\ref{eq:dC2 map}) which
is a quotient of two quadratic polynomials in $m_k,p_k$.

If the coefficients $b_k, c_k$ are as in eqs. (\ref{eq: Clebsch 2
b}), (\ref{eq: Clebsch 2 c}), then the integral $L$ can be taken
as
\begin{equation} \nonumber
L = \frac{E_1(\omega_1
m_1^2+\omega_2\omega_3p_1^2)+E_2(\omega_2m_2^2+\omega_3\omega_1p_2^2)+
E_3(\omega_3m_3^2+\omega_1\omega_2p_3^2)}
{1+\epsilon^2\omega_1(\omega_1 m_1^2
+\omega_2\omega_3p_1^2)+\epsilon^2 \omega_2(\omega_2 m_2^2
+\omega_3\omega_1p_2^2)+\epsilon^2\omega_3(\omega_3 m_3^2
+\omega_1\omega_2p_3^2)}\,,
\end{equation}
with
\begin{equation}\label{eq: E}
E_1=\omega_3\omega_1+\omega_1\omega_2-\omega_2\omega_3, \quad
E_2=\omega_1\omega_2+\omega_2\omega_3-\omega_3\omega_1, \quad
E_3=\omega_2\omega_3+\omega_3\omega_1-\omega_1\omega_2.
\end{equation}
In this case there holds:
\begin{eqnarray*}
\frac{e_1}{\omega_2\omega_3}=\frac{e_4}{\omega_1} & = &
   \frac{E_1}{L}-\epsilon^2\omega_1\\
& = &
\frac{E_1+\epsilon^2(\omega_1-\omega_2)E_3(\omega_2m_2^2+\omega_3\omega_1p_2^2)+
          \epsilon^2(\omega_1-\omega_3)E_2(\omega_3m_3^2+\omega_1\omega_2p_3^2)}
{E_1(\omega_1m_1^2+\omega_2\omega_3p_1^2)+
 E_2(\omega_2m_2^2+\omega_3\omega_1p_2^2)+
 E_3(\omega_3m_3^2+\omega_1\omega_2p_3^2)},\\
 \\
\frac{e_2}{\omega_3\omega_1}=\frac{e_5}{\omega_2} & = &
  \frac{E_2}{L}-\epsilon^2\omega_2\\
& = &
\frac{E_2+\epsilon^2(\omega_2-\omega_3)E_1(\omega_3m_3^2+\omega_1\omega_2p_3^2)+
          \epsilon^2(\omega_2-\omega_1)E_3(\omega_1m_1^2+\omega_2\omega_3p_1^2)}
{E_1(\omega_1m_1^2+\omega_2\omega_3p_1^2)+
 E_2(\omega_2m_2^2+\omega_3\omega_1p_2^2)+
 E_3(\omega_3m_3^2+\omega_1\omega_2p_3^2)},\\
\frac{e_3}{\omega_1\omega_2}=\frac{e_6}{\omega_3} & = &
  \frac{E_3}{L}-\epsilon^2\omega_3\\
& = &
\frac{E_3+\epsilon^2(\omega_3-\omega_1)E_2(\omega_1m_1^2+\omega_2\omega_3p_1^2)+
          \epsilon^2(\omega_3-\omega_2)E_1(\omega_2m_2^2+\omega_3\omega_1p_2^2)}
{E_1(\omega_1m_1^2+\omega_2\omega_3p_1^2)+
 E_2(\omega_2m_2^2+\omega_3\omega_1p_2^2)+
 E_3(\omega_3m_3^2+\omega_1\omega_2p_3^2)}.
\end{eqnarray*}
The numerator of $L$,
\[
L(m,p,0)=E_1(\omega_1m_1^2+\omega_2\omega_3p_1^2)+
E_2(\omega_2m_2^2+\omega_3\omega_1p_2^2)+
E_3(\omega_3m_3^2+\omega_1\omega_2p_3^2),
\]
is a linear combination of quadratic integrals of motion of the
continuous Clebsch system.
\end{theorem}
{\bf Proof.} We will only present the proof for the second Clebsch
flow. The claim of the theorem refers to the linear system
\begin{equation}\nonumber
(e_1 p_1^2 + e_2 p_2^2 + e_3 p_3^2 + e_4 m_1^2
+ e_5 m_2^2 + e_6 m_3^2) \circ f^i(m,p) = 1
\end{equation}
for $i$ from the ranges containing 6 consecutive numbers, such as
$i\in[-2,3]$ or $i\in[-3,2]$.  As the solution of such a system
clearly requires more iterates of the map $f$ than could be
handled symbolically, we follow recipe (E) and look for linear
relations between $e_i$. It turns out to be possible to identify
the following five relations:
\begin{eqnarray}
 \omega_1e_1 - \omega_2\omega_3e_4 &=& 0, \label{eq: dCS2 linrel 1}\\
 \omega_2e_2 - \omega_3\omega_1e_5 &=& 0, \label{eq: dCS2 linrel 2}\\
 \omega_3e_3 - \omega_1\omega_2e_6 &=& 0, \label{eq: dCS2 linrel 3}\\
 e_1-e_2-(\omega_1-\omega_2)e_6 &=&
\epsilon^2\omega_3^2(\omega_1-\omega_2), \label{eq: dCS2 linrel 4}\\
 e_3-e_1-(\omega_3-\omega_1)e_5 &=&
\epsilon^2\omega_2^2(\omega_3-\omega_1). \label{eq: dCS2 linrel 5}
\end{eqnarray}
Of course, there holds also a third non-homogeneous relation:
\begin{equation}\nonumber
e_2-e_3-(\omega_2-\omega_3)e_4 =
\epsilon^2\omega_1^2(\omega_2-\omega_3),
\end{equation}
but actually it is a consequence of the previous five. As usual,
these (at this point conjectural) identities can be (and have
been) found using the PSLQ algorithm. Now we obtain the six
functions $e_i$ by solving a simple system of six linear equations
which involves no iterates of the map $f$ at all and consists of
\[
e_1 p_1^2 + e_2 p_2^2 + e_3 p_3^2 + e_4 m_1^2 + e_5 m_2^2 + e_6
m_3^2  =  1\,
\]
along with the relations (\ref{eq: dCS2 linrel 1})--(\ref{eq: dCS2
linrel 5}). The solution is given in the formulation of the
theorem. To prove that the function $L$ is an integral of motion
one can use a straightforward computation using MAPLE. Also a
proof based on the equations of motion alone can be given, similar
to the proof for $L$ (see proof of Theorem \ref{Th: dC basis 1}).
The last claim of the theorem about $L(x,0)$ follows in the limit
$\epsilon\to 0$, but can be also easily checked directly, by
verifying conditions (\ref{eq: c thru b}) for
$b_i=\omega_j\omega_kE_j$ and $c_i=\omega_iE_i$ with $E_i$ from
eq. (\ref{eq: E}). These conditions are satisfied due to the
identities
\[
\omega_jE_i-\omega_iE_j=(\omega_i-\omega_j)E_k,
\]
where $(i,j,k)$ is any permutation of (1,2,3). \hfill $\Box$

\section{Conclusions}
\label{Sect: conclusions}

We established the integrability of the Hirota-Kimura-type
discretization of the Clebsch system, in the sense of
\begin{itemize}
 \item existence, for every initial point
$(m,p)\in\bbR^6$, of a four-dimensional pencil of quadrics
containing the orbit of this point; in our terminology, this can
be formulated as existence of a HK-basis with a four-dimensional
null-space, consisting of quadratic monomials;
 \item existence of four functionally independent integrals of
motion (conserved quantities).
\end{itemize}
Numerical experiments show that this remains true also for an
arbitrary flow of the Clebsch system. It is interesting to remark
that the maps generated by Hirota-Kimura discretizations of
various flows do not commute with each other. It would be
important to understand whether some analog of commutativity of
the continuous flows survives in the discrete situation.

Our investigations were based mainly on computer experiments. Our
proofs are computer assisted and were obtained with the help of
symbolic calculations with MAPLE, SINGULAR and FORM. A general
structure behind these facts, which would provide us with more
systematic and less computational proofs and with more insight,
remains unknown. In particular, nothing like a Lax representation
has been found. Nothing is known about the existence of an
invariant Poisson structure for these maps. (For a simpler system,
Hirota-Kimura discretization of the Euler top, an invariant volume
measure as well as a bi-Hamiltonian structure have been found in
\cite{PS}.)

Hirota and Kimura demonstrated that their discretization leads to
an integrable map also for the Lagrange top \cite{KH}. Our
preliminary investigations have shown remarkable features pointing
towards the integrability of the Hirota-Kimura discretizations of
the following systems: Zhukovsky-Volterra gyrostat; $so(4)$ Euler
top and its commuting flows; Volterra and Toda lattices; classical
Gaudin magnet. Based on these observations, we formulate the
following hypothesis.

\begin{conjecture}
For any algebraically completely integrable system with a
quadratic vector field, its Hirota-Kimura discretization remains
algebraically completely integrable.
\end{conjecture}

If true, this statement could be related to addition theorems for
multi-dimensional theta-functions. Such a relation has been
already established for the Hirota-Kimura discretization of the
Euler top, which can be solved explicitly in elliptic functions
\cite{theta}. In our ongoing investigations, we hope to establish
integrability of the above mentioned discrete time systems and to
uncover general mechanisms behind it.

\section*{Aknowledgments}

M.P. was partially supported by the European Community through the FP6 Marie
Curie RTN ENIGMA (Contract number MRTN-CT-2004-5652) and by
the European Science Foundation project MISGAM. M.P. acknowledges the
hospitality of the Mathematics Department of the Technical University of
Munich, where the present work was finalised.


\begin{thebibliography}{}
\bibitem[Worksheets]{WS}
Maple worksheets for this work are available on request from {\tt
suris}@{\tt ma.tum.de}
\end{thebibliography}
\end{document}